\documentclass[useAMS,usenatbib]{mn2e}
\pdfoutput=1
\usepackage{amssymb}
\usepackage{amsmath}
\usepackage{natbib}
\usepackage[pdftex]{graphicx}
\usepackage{subfigure}
\usepackage{booktabs}

\title[On the tidal evolution of Hot Jupiters on inclined orbits]
      {On the tidal evolution of Hot Jupiters on inclined orbits}
      \author[A.J. Barker \& G.I. Ogilvie]{Adrian J. Barker\thanks{E-mail:
	  ajb268@cam.ac.uk} and Gordon I. Ogilvie \\
	Department of Applied Mathematics and Theoretical
	Physics, University of Cambridge, Centre for Mathematical Sciences, \\ Wilberforce Road,
	Cambridge CB3 0WA, UK}
      
\begin{document}
	
\date{Accepted 2009 February 25.  Received 2009 February 24; in original form 2009
January 16}

\pagerange{\pageref{firstpage}--\pageref{lastpage}} \pubyear{2009}

\maketitle

\label{firstpage}

\begin{abstract}
Tidal friction is thought to be important in determining the long--term 
spin--orbit evolution of short--period extrasolar planetary systems. 
Using a simple model of the orbit--averaged effects of tidal friction 
\citep{Eggleton1998}, we study the evolution of close--in planets
on inclined orbits, due to tides.
We analyse the 
effects of the inclusion of stellar magnetic braking by performing 
a phase--plane analysis of a simplified system of equations,
including the braking torque. The inclusion of magnetic braking is found to be
important, and its neglect can result in a very different system
history. We then present the results of numerical integrations of the
tidal evolution equations, where we find that
it is essential to consider coupled evolution of the orbital and
rotational elements, including dissipation in both the star and
planet, to accurately model the evolution. 

The main result of our integrations is that for typical Hot Jupiters,
tidal friction aligns the stellar spin with the orbit on a similar
time as it causes the orbit to decay. This tells us that if a planet is observed to be aligned,
then it probably formed coplanar. This reinforces the
importance of Rossiter--McLaughlin effect observations in
determining the degree of spin--orbit alignment in transiting
systems.

We apply these results to the only 
observed system with a spin--orbit misalignment, XO-3, and constrain
the efficiency of tidal dissipation (i.e. the modified tidal quality factors $Q^{\prime}$) in both the star and 
the planet in this system. Using a model in which inertial waves are excited by
tidal forcing in the outer convective envelope and dissipated by turbulent
viscosity, we calculate $Q^{\prime}$ for a range of F--star models, and 
find it to vary considerably within this class of stars.
This means that using a
single $Q^{\prime}$, and assuming that it applies to all stars, is
probably incorrect. In addition, we propose an explanation for the
survival of two of the planets on the tightest orbits, WASP-12 b and
OGLE-TR-56 b, in terms of weak dissipation in the star, as a result of
their internal structures and slow rotation periods.

\end{abstract}


\section{Introduction}
Since the discovery of the first extrasolar planet around a solar--type
star \citep{MQ1995}, observers have now detected more than 300
planets\footnote{see http://exoplanet.eu/ for the latest updates}
around stars outside the solar system. Many of these planets have
roughly Jovian masses and orbit their host stars in orbits with semi-major
axes less than $0.1$ AU, the so--called ``Hot Jupiters'' (HJs). In both
of the giant planet formation scenarios (see
\citealt{Papaloizou2006} for a comprehensive review),
core accretion and gravitational instability, it is difficult to produce
HJs in situ. Close--in
planets are likely to form in colder regions of the protoplanetary disc, much further out ($a \sim $
several AU), before a migratory process brings the planet in
towards the star and to its present location \citep{LBR1996}. 

The formation of systems of giant planets can be thought of as
occurring in two oversimplified stages
\citep{JT2008}. During stage 1 the cores of the giant planets are formed, they accrete
gas and undergo migration, driven by the dynamical interaction between the
planets and the gaseous protoplanetary disc (see \citealt{PapProt2007} for a
recent review).  This stage lasts a few Myr until the gas dissipates, by which
time a population of gas giants may exist.  If these form sufficiently
closely packed then stage 2 follows (some evidence in
favour of such packing in multiple--planet systems is given by \citealt{BarnesGreenberg2007}).
This stage lasts from when the disc has dissipated until the
present, and primarily involves gravitational interactions and
collisions between the planets. Recent studies of stage 2
(\citealt{JT2008}; \citealt{Chatterjee2008}; \citealt{FordRasio2008}) have shown that this is a chaotic era, in
which planet--planet scatterings force the
ejection of all but a few ($\sim 2-3$) planets from the system, in a
period of large--scale dynamical instability lasting $\la
10^{8}$yr. This mechanism can excite the eccentricities of the planets
to levels required to explain observations. 

Planet--planet scatterings tend also to excite the inclinations of the 
planets with respect to the initial symmetry plane of the system
\citep{JT2008}, though this has been found to be less efficient than the excitation of eccentricity. This 
potentially leads to observable
consequences via the Rossiter--McLaughlin (RM) effect 
(\citealt{Rossiter1924}; \citealt{Mclaughlin1924};
\citealt{Winn2005}).

The RM effect is a spectral distortion of the radial velocity
data that results from the planet occulting a spot on
the rotating surface of the star as it transits the stellar
disc. High--precision radial velocity data during a
transit allow a determination of the angle ($\lambda$) between the 
sky--projected angular momentum vectors of the planetary orbit ($\mathbf{h}$)
and stellar spin ($\mathbf{\Omega}$), through the RM effect. This measured value
$\lambda$ is not necessarily the same as the inclination (or stellar
obliquity) $i$, which is the angle between the equatorial plane of the star and
orbital plane of the planet (defined by
$\cos i = \hat{\mathbf{\Omega}} \cdot \hat{\mathbf{h}}$), since $\lambda$ is
just a sky--projection of this angle. Nevertheless, $\lambda$ gives a
lower bound on the angle between these two vectors, and they are related by
$\cos i = \cos I_{\star} \cos I_{p} + \sin I_{\star} \sin I_{p} \cos
\lambda$, where $I_{p}$ and $I_{\star}$ are the angles of inclination of the planetary
orbital plane and stellar equatorial plane, to the plane of the sky. For a transit,
the orbit must be close to edge--on, so
$I_{p}\sim 90^{\circ}$, giving $\cos i \simeq \sin I_{\star}\cos \lambda$.

The RM effect has now been used to measure the degree of 
spin--orbit alignment in 11 systems \citep{Winn2008}, though this number is
expected to grow rapidly over the next few years. These systems are 
currently all consistent with $\lambda$ being
zero, with the exception of XO-3 b \citep{Hebrard2008}, which is
discussed in \S \ref{XO-3} below.

It is possible for HJs orbiting a host star which has a distant and inclined
stellar companion, or massive inclined outer planetary companion, to undergo another 
type of migration. This is Kozai migration \citep{WuMurray2003}.
The presence of such an outer companion to an exoplanet host star could
cause Kozai oscillations, which produce periods of extreme
eccentricity in the planetary orbit, if various conditions are satisfied
(e.g. see section 1.2 of \citealt{Fabrycky2007}). The subsequent tidal
dissipation that occurs during the periods of small pericentre distance leads to
gradual inward migration of the planet. It has
even been proposed that a combination of planet--planet scattering, tidal
circularisation and the Kozai mechanism using outer planets, can produce
HJs around single stars \citep{Nagasawa2008}. HJs produced from these processes
generally have their orbital angular momentum vector misaligned with respect
to the stellar spin axis by large angles -- occasionally larger than $90^{\circ}$ 
(\citealt{Fabrycky2007}; \citealt{Nagasawa2008}). 

Misaligned orbits are not predicted from stage 1
alone, so if $\lambda$ is measured
to be appreciably nonzero in enough systems, then it could be seen as
evidence for planet--planet
scattering or Kozai migration. This is because 
gas--disc migration does not seem able to excite orbital inclination
(\citealt{LubowOgilvie2001}; \citealt{Cresswell2007}). Alternatively,
if observed planets are all found with
$\lambda$ consistent with zero, this could rule out planet--planet scattering
or Kozai migration as being
of any importance.

One important consideration is that at such close
proximity to their parent stars, strong tidal interactions between the
star and planet are expected to cause significant long--term spin--orbit
evolution, including changes to the value of
$\lambda$ (actually the true spin--orbit
misalignment angle $i$) over time. If tides can change
$\lambda$ since the time of formation, then we may have difficulty in
distinguishing migration caused by planet--planet scattering and Kozai
oscillations, from gas--disc migration. The tidal evolution of such
inclined orbits must therefore be considered an important goal in planetary evolution
studies. In this paper we approach the problem of studying the effects of
tidal friction on such inclined orbits.

\section{Previous work on tidal evolution relevant to close--in planets}

There has been much recent interest into the subject of tidal evolution, and our
work can be considered to follow on from some of the more recent
developments in the field. \cite{WS2002} studied the 
effects of stellar spin--down on
tidal evolution. They developed a sophisticated model including the
effects of stellar evolution, resonance
locking and magnetic braking. They studied
dynamical tides in
solar--type stars, and found that the spin-down due to magnetic braking
causes resonance locking to become more intense, which leads to faster
orbital decay of a planet. 

\cite{DobbsLin2004} also studied
the effects of stellar spin--down, though they considered equilibrium
tides, and we adopt the same model of tidal friction. 
They studied tidal eccentricity evolution, and
proposed an explanation for the
coexistence of both circular and eccentric orbits for the planets in the
period range 7-21 days, the so--called borderline planets, as
the result of the variation in spin--down rates of young
stars. Planets with orbital periods less than 6 days are sufficiently
close to their host stars for tidal dissipation in the planet to be mostly
able to account for their negligible
eccentricities. Those with periods longer than 21 days are
negligibly affected by tides and have been observed with 
a range of eccentricities, so the borderline planets
are referred to as such since they lie in the crossover
between these two regimes. Both of these works highlight the 
importance of studying the effects of stellar
spin--down on tidal evolution.

Recently \cite{Pont2008} discussed empirical evidence for tidal
spin--up of exoplanet host stars, and found indications 
of such a process occurring in the present sample of transiting
planets. He also
proposed that the mass--period relation
of close--in planets could be
accounted for by tidal transfer of angular momentum from the orbit to
the spin of the star. We must note, however, that
angular momentum losses from the system due to magnetic braking are
likely to change this picture considerably. In particular, the
attainment of a spin--orbit synchronous state becomes questionable when
angular momentum is being removed from the system -- this is
discussed in \S \ref{phaseplaneanalysis} below. We note that
his conclusions also depend on assumptions regarding the stellar $Q^{\prime}$ -- see \S
\ref{introtide}.

\cite{JacksonI2008} found that simple timescale
considerations of circularisation may not accurately represent the
true evolution from considering coupled evolution of the eccentricity
and semi-major axis. In addition, they found that it is inaccurate to neglect the
combined effects of the stellar and planetary tides in computing the
evolution.

Here we aim to study the accuracy of the simple
timescale estimates for tidal evolution, when coupled evolution of the
orbital and rotational elements is considered, in a more general model
of the long--term effects of tidal friction than \cite{JacksonI2008}
consider. We include stellar spin--down, and study its effects in a simplified
system. In particular, we investigate the tidal evolution of inclination,
since this has not been done in previous studies.

\section{Model of Tidal Friction adopted}
\subsection{General introduction to tidal friction}
\label{introtide}
The tidal interaction between two orbiting bodies acts to continually
change the orbital and rotational system parameters, and continually
dissipates energy. Ultimately -- in
the absence of angular momentum loss from the system --
either an equilibrium state is asymptotically approached, or the two
bodies spiral towards each other at an accelerating rate, and
eventually collide. The equilibrium state is characterised by
coplanarity (the equatorial planes of the bodies coincide with the
orbital plane), circularity and corotation (rotational frequencies of each
body match the orbital frequency) \citep{Hut1980}.

The efficiency of tidal dissipation in a body is often parametrised by a
dimensionless quality factor $Q$, which reflects the fact the body
undergoes a forced oscillation and dissipates a small fraction of the
associated energy during each oscillation period. This is analogous to
the quality factor in a forced, damped harmonic oscillator
\citep{MurrayDermott1999}, and is defined by
\begin{eqnarray*}Q = 2\pi E_{0} \left(\oint -\dot E \,\mathrm{d}t\right)^{-1}, \end{eqnarray*}
where $E_{0}$ is the maximum energy stored in an oscillation and the
integral represents the energy dissipated over one cycle. This is
related to the time lag $\tau$ in the response of the body to tidal forcing
of frequency $\hat \omega$ by $Q^{-1}=\hat \omega \tau$, when $Q \gg 1$. We find it
convenient to define $Q^{\prime} = \frac{3Q}{2k}$,
where $k$ is the second--order potential Love number of the body,
since this combination always
appears together in the evolutionary equations. $Q^{\prime}$ reduces to
$Q$ for a homogeneous fluid body, where $k=\frac{3}{2}$.

The problem of determining the efficiency of tidal dissipation, and therefore
quantifying the evolution of the system, amounts to calculating $Q^{\prime}$
factors for each body. $Q^{\prime}$ is a function of the 
tidal frequency (and possibly the 
amplitude of the tidal disturbance), being the
result of complex dissipative processes in each body
\citep{Zahn2008}. For rotating fluid bodies,
such as giant planets and stars, calculations of the excitation 
and dissipation of internal waves have
indicated that $Q^{\prime}$ varies in a complicated way with the tidal frequency (see
\citealt{Savonije1995}; \citealt{Savonije1997};
\citealt{Papaloizou1997}; \citealt{Gio2004}, hereafter OL04;
\citealt{Gio2007}, hereafter OL07).  These
calculations rely on a variety of approximations and involve uncertainties,
particularly regarding the interaction of the waves with convection.

Typically assumed values of $Q^{\prime} \sim 10^{6}$ for stars 
are roughly consistent with observational data regarding the circularisation
periods of binary stars (OL07). In addition, the magnitude of $Q$
for HJs is often assumed to be similar to that for Jupiter, which has
been inferred to be in the range $6\times 10^{4} - 2\times 10^{6}$
\citep{YoderPeale1981}. This estimate is based on a model
of the tidal origin of the Laplace resonance among the Galilean satellites;
however, it has been argued that even if the origin of the resonance is
primordial, the average $Q$ cannot be far from these bounds \citep{Peale2002} -
giving $Q^{\prime} \sim 10^{6}$ (since $k \simeq 0.38$ for Jupiter).
This estimate also appears consistent with the work of
\cite{JacksonI2008}, who found that one can
reproduce the outer planet ($a>0.2\,\mathrm{AU}$) eccentricity
distribution from integrating the tidal evolution equations backwards
in time for the observed close--in planets ($a < 0.2$ AU) quite well 
if $Q \sim 10^{5.5}$ for stars and $Q \sim 10^{6.5}$ for HJs. However, 
this stellar $Q^{\prime}$ is difficult to reconcile with the existence of the planets
on the tightest orbits, such as WASP-12 b \citep{Hebb2008} and OGLE-TR-56
b \citep{Sasselov2003}, since it would imply that the inspiral time for these
planets would be much less than the age of the system. That several planets
have been found with similarly short periods makes this seem unlikely on
probabilistic grounds. Additionally, \cite{JacksonI2008} assume
that $Q^{\prime}$ is the same for all exoplanet host stars; this may 
be an oversimplification, as is discussed in \S \ref{FstarsQ} below.

It would seem more plausible that the relevant
stellar $Q^{\prime}$ is in fact higher than this, and this is supported by the
theoretical work of OL07. They also propose that the efficiency of tidal dissipation in solar--type stars may be
different when the orbiting companion is a $\sim M_{\odot}$ star
in a close binary, than when its orbiting companion is a close--in gas giant
planet. This could be a result of differences in the tidal and spin frequencies in the
two situations, which may allow the excitation of inertial waves in the
convective envelope of the star in the former case but usually not in the
latter. It could also be a result of the
excitation of 
internal inertia--gravity modes at the boundary between the convective and radiative
zones, and their resulting dissipation in the radiative core of
solar--type stars. If these waves are of small amplitude, then they are
unlikely to achieve sufficient nonlinearity to prevent coherent
reflection from the centre of the star, hence they form global modes,
and only weakly contribute to the dissipation \citep{GoodmanDickson1998}.
If these waves achieve sufficient nonlinearity, then they could overturn 
the local entropy stratification. This may result in wave breaking,
which could dissipate energy in the wave, or in incoherent reflection of the waves
from the centre of the star, preventing the formation
of global modes, and enhancing the dissipation. Simple estimates of
when these waves become nonlinear 
indicate that the waves excited by HJs marginally achieve
sufficient nonlinearity to
disrupt the reflection of the waves. This is in contrast to the close binary
circularisation problem, in which these waves are likely always to
achieve sufficient nonlinearity. This may result in much higher $Q^{\prime}$
values relevant to the survival of the very close--in planets,
potentially explaining
the discrepancy between the circularisation of binary stars and the
survival of the very close--in planets.

\subsection{Model adopted -- see Appendix \ref{Appendix} for details}

In light
of the uncertainties involved in calculating $Q^{\prime}$, and the difficulty
of calculating the evolution when $Q^{\prime}$ is a complicated function of $\hat \omega$, we adopt a
simplified model, based on a frequency--independent lag time --
though see discussion in Appendix \ref{Appendix}. We adopt the model of
\cite{Eggleton1998}, which is based on the equilibrium tide model of
\cite{Hut1981}. In this formulation, we calculate the evolution of the 
specific angular momentum of the planetary orbit \begin{eqnarray*}\mathbf{h} = \mathbf{r} \times \dot{\mathbf{r}} = n a^{2}
\sqrt{1-e^{2}} \; \hat{\mathbf{h}},\end{eqnarray*} together
with its eccentricity vector $\mathbf{e}$, and the stellar and planetary
spin vectors $\mathbf{\Omega}_{1}$ and $\mathbf{\Omega}_{2}$. The eccentricity vector
has the magnitude of the eccentricity, and points in the direction of
periastron, and is defined by 
\begin{eqnarray*} \mathbf{e} = \frac{\dot{\mathbf{r}} \times \mathbf{h}}{Gm_{12}} - \hat{\mathbf{
    r}},
\end{eqnarray*} where $m_{12}=m_{1}+m_{2}$ is the sum of the stellar
and planetary masses. 

Both $\mathbf{h}$ and $\mathbf{e}$ are conserved for
an unperturbed Keplerian orbit; therefore under weak external
perturbations their components vary slowly compared with the orbital
period. This allows averaging of the effects of the tidal perturbation over
a Keplerian orbit, resulting in a set of secular evolution equations
for the rotational and orbital elements. 

This formulation is
beneficial because it can treat arbitrary orbital eccentricities
and stellar and planetary obliquities, unlike other models which are
only valid to a given order in the eccentricity, or for small (or
zero) orbital inclinations (\citealt{GoldSot1966};
\citealt{JacksonI2008}; \citealt{Hut1981}). Using the secular
evolution equations allows us to perform integrations quickly that
represent dynamical evolution over billions of years. The full set of equations
is presented in Appendix \ref{Appendix} in a form which is straightforward to
numerically integrate. These
equations have been written in such a way as to 
eliminate references to the basis vectors chosen in their
representation, since the eccentricity basis vector is undefined for a
circular orbit. In this
form, the equations are regular at $e=0$, unlike those in
\cite{Eggleton2001} and \cite{ML2002}.

\section{Magnetic braking}

Observations of solar--type stars have shown that the mean 
stellar rotational velocity decreases with time
\citep{Skumanich1972}, 
following the relation $\Omega \propto t^{-1/2}$, where $t$ is
the main--sequence age. 
This is the empirical Skumanich relation, and can be
interpreted as telling us that solar--type stars have been undergoing continuous
spin--down since they first started on the main sequence.
Magnetic braking by a magnetised outflowing wind has long been
recognised as an important mechanism for the removal of angular
momentum from rotating stars \citep{WD1967}, and such a mechanism
seems able to explain most of the
observed stellar spin--down \citep{Barnes2003}.

Although the Skumanich law is well established for stars with
rotational velocities in the range $1-30 \; \mathrm{km}\,\mathrm{s}^{-1}$, it overestimates
the spin--down rates of stars up to $t \sim 10^{8}$ yrs, and thus
cannot explain the presence of fast rotators in the Pleiades \citep{IvanovaTaam2003}. 
As a resolution to this problem it has been suggested that the angular
momentum loss rate for high rotation rates could be reduced, as a
result of the saturation of the stellar dynamo
\citep{MacBren1991}, or alternatively due to a reduction in the
number of open field lines in a complex magnetic field topology
\citep{TaamSpruit1989}. These and similar approaches lead
to modified models of the magnetic braking torque for fast rotators,
and several such models have been proposed (e.g.\ \citealt{IvanovaTaam2003};
\citealt{Holzwarth2005}). Nevertheless, \cite{Barnes2003} finds that the
Skumanich relation is remarkably accurate at modelling the spin--down
of Sun--like stars that are not rapid rotators, so to a
first approximation, a magnetic braking torque based on the empirical
Skumanich law is best for our purposes.

Here we include the effects of magnetic braking in
the tidal evolution equations, through the inclusion of 
the \cite{VerbuntZwaan1981} braking torque, with the particular
coefficients of \cite{DobbsLin2004}, as follows (see Appendix~\ref{Appendix})
\begin{eqnarray}
  \label{MBtorque}
  \dot {\bmath\omega}_{\mathrm{mb}} = - \alpha_{\mathrm{mb}} \; \Omega_{1}^{2} \;
  \mathbf{\Omega}_{1},
\end{eqnarray}
where $\alpha_{\mathrm{mb}} = 1.5 \times 10^{-14}\gamma$ yrs. $\gamma$ is a correction
factor for F-dwarfs, which takes the value 0.1 for an F dwarf, but is
unity for a G or K dwarf. We can also define a magnetic braking
timescale $ \tau_{\mathrm{mb}} \equiv \frac{\Omega_{1}}{\dot \omega_{\mathrm{mb}}} =  
\frac{1}{\alpha_{\mathrm{mb}}}\frac{1}{\Omega_{1}^{2}}$,
which is $\sim 10^{10}$ yrs for the Sun.

\section{Analysis of the effects of magnetic braking on tidal evolution for a
  simplified system}
\label{phaseplaneanalysis}
\subsection{Circular, coplanar orbit with magnetic braking}
We first study the effects of magnetic braking on a simplified system of a
circular, coplanar orbit under the influence of only the
tide that is raised on the star by the planet,
and magnetic braking. We have neglected the tide in the planet
here since the
moment of inertia of the planet is much smaller than that of star and
the orbit (i.e. $I_{2} \ll I_{1} \sim m_{2}a^{2}$), so to a first
aproximation we can neglect the effects of planetary spin; in any case the
planetary spin is expected to synchronize rapidly with the orbit. The following set of
dimensionless equations can be derived from the full set of equations
in Appendix \ref{Appendix}:
\begin{eqnarray}
\label{eqn:ODE1}
\frac{d\tilde \Omega}{d\tilde t} &=&
\tilde n^{4}\left(1-\frac{\tilde \Omega}{\tilde n}\right) -
A \: \tilde \Omega^{3}, \\
\frac{d\tilde n}{d\tilde t} &=& 3 \: \tilde n^{\frac{16}{3}}\left(1-\frac{\tilde \Omega}{\tilde n}\right),
\label{eqn:ODE2}
\end{eqnarray}
where we have normalised the stellar spin frequency $\Omega_{1}$ and
orbital mean motion $n$ to the orbital frequency at the
stellar surface, together with a factor $C^{3/4}$. $C$ is the ratio of
the orbital angular momentum of a mass $m_{2}$ in an orbit with semi-major
axis equal to the stellar radius $R_{1}$, to
the spin angular momentum of an equally rapidly rotating star of radius
$R_{1}$, mass $m_{1}$ and dimensionless radius of gyration
$r_{g1}$. The reduced mass is
$\mu=\frac{m_{1}m_{2}}{m_{1}+m_{2}}$. $C$ is important for
classifying the stability of the equilibrium curve $\tilde \Omega = \tilde
n$ in the absence of magnetic braking, and it can be shown from energy
and angular momentum considerations that this
equilibrium is stable if $\tilde n \leq 3^{-\frac{3}{4}}$ -- equivalent to the
statement that no more than a quarter of the total angular
momentum can be in the form of spin angular momentum for stability \citep{Hut1980}. We have thus defined the
following dimensionless quantities:

\begin{eqnarray*}
  \tilde \Omega &=& \Omega_{1}\left(\frac{R_{1}^{3}}{Gm_{12}}
  \right)^{\frac{1}{2}}C^{-\frac{3}{4}}, \\
  \tilde n &=& n\left(\frac{R_{1}^{3}}{Gm_{12}}
  \right)^{\frac{1}{2}}C^{-\frac{3}{4}}, \\
  C &=& \frac{\mu R_{1}^{2}}{I_{1}} = \frac{\mu}{r_{g 1}^{2} \: m_{1}}, \\
  \tilde t &=& \left(\frac{Gm_{12}}{R_{1}^{3}}\right)^{\frac{1}{2}} \: 
    \left(\frac{9}{2Q^{\prime}_{1}}\right)\left(\frac{m_{2}}{m_{1}}\right) 
    \: C^{\frac{13}{4}} \; t, \\
    A &=& \alpha_{\mathrm{mb}} \:
    \: \left(\frac{Gm_{12}}{R_{1}^{3}}\right)^{\frac{1}{2}} \:
    \left(\frac{2 Q^{\prime}_{1}}{9}\right)\left(\frac{m_{1}}{m_{2}}\right)
    C^{-\frac{7}{4}}.
\end{eqnarray*}

There is only one parameter ($A$) that completely characterises
the solution in the $(\tilde n,\tilde \Omega)$--plane, and its value
may be estimated as 
\begin{eqnarray*}
    A \simeq 100 \: \gamma
    \left(\frac{Q_{1}^{\prime}}{10^{6}}\right),
\end{eqnarray*} for a Jupiter--mass planet orbiting a Sun--like star undergoing
magnetic braking (with standard $\alpha_{\mathrm{mb}}$ and with $Q^{\prime} = 10^{6}$).
The size of this term shows that in general magnetic braking dominates
the stellar spin evolution. Note that in
the absence of magnetic braking ($A=0$), Eqs.~\ref{eqn:ODE1} and
\ref{eqn:ODE2} do not contain reference to the masses of the star and
planet or to the tidal $Q^{\prime}$ of the star. The parameter $A$, together with the initial conditions ($\tilde
n_{0},\tilde \Omega_{0}$), completely determines the evolution.

\begin{figure}
  \centering
  \subfigure{\label{Figmb1}\includegraphics
    [width=0.495\textwidth]{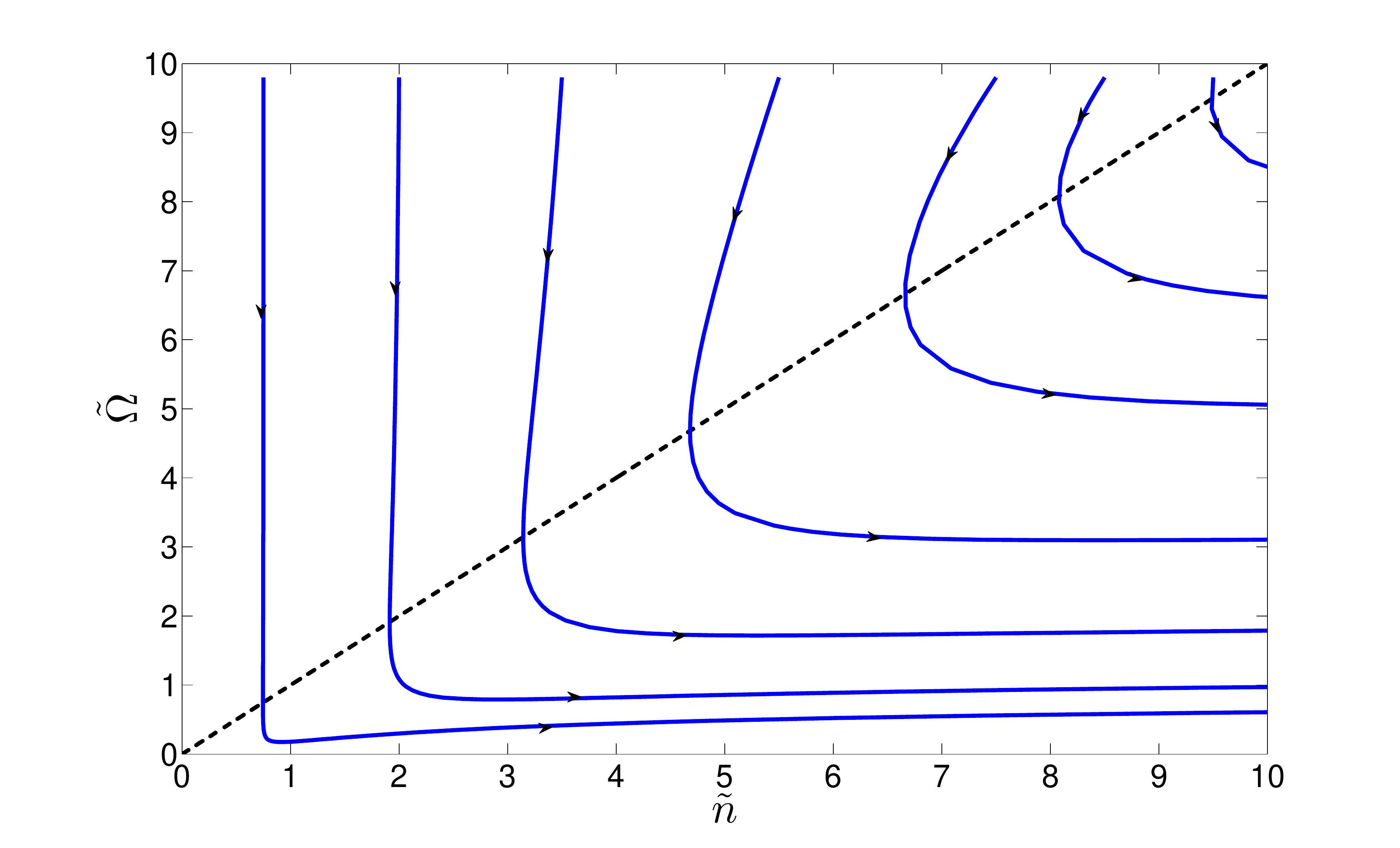}}                \\
  \subfigure{\label{Figmb2}\includegraphics[width =
      0.495\textwidth]{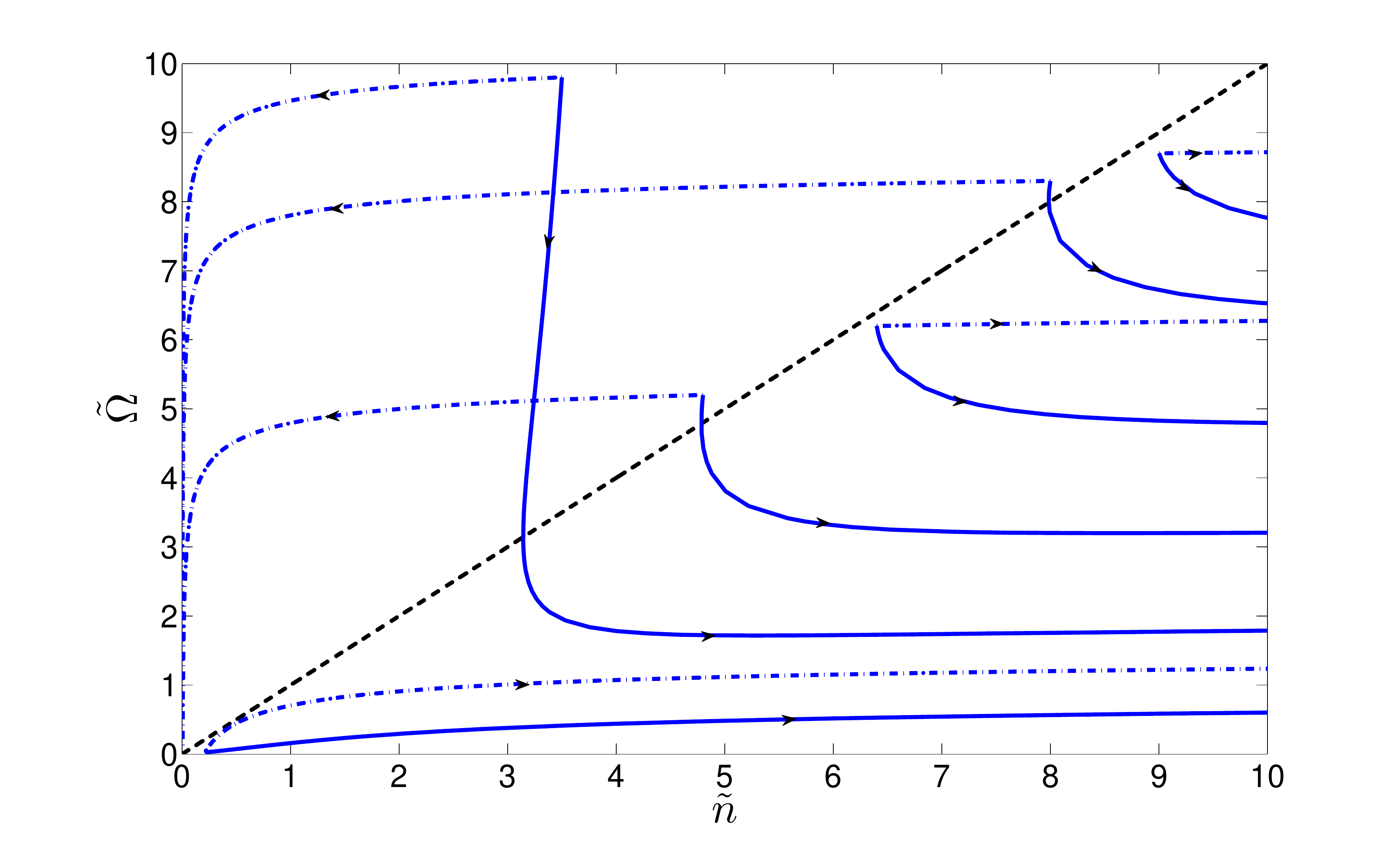}}
  \caption{($\tilde n,\tilde \Omega$)--plane with $A = 100$
    for a HJ orbiting a Sun--like star. The diagonal dashed line in each
    plot corresponds to corotation ($\tilde \Omega =\tilde n$). 
    Top: magnetic braking spins the star down so that the planet finds
    itself inside corotation, where the sign of the tidal torque
    changes, and planet is subject to tidally induced orbital decay. For an initially high
    $\tilde n$ outside corotation tidal friction efficiently transfers angular momentum from spin to orbit,
    which pushes the planet outwards.
    Bottom: Solutions with the same initial
    conditions are plotted with and without magnetic braking for a HJ around a solar--type star, with
    dot-dashed lines having $A=0$ and solid lines have $A = 100$. The dot--dashed lines
    are also curves of constant total angular momentum. This shows that the
    inclusion of magnetic braking is extremely important in
    determining the secular evolution of the system, and its absence
    results in a very different evolutionary history unless $\tilde
    \Omega \ll \tilde n$ in the initial state.}
  \label{fig:withMBplots}
\end{figure}

We plot some solutions on the $(\tilde n,\tilde
\Omega)$--plane in Fig.~\ref{fig:withMBplots} by integrating
Eqs.~\ref{eqn:ODE1} and \ref{eqn:ODE2} for
various initial conditions. Fig.~\ref{fig:withMBplots} shows two phase 
portrait plots, which show the general qualitative behaviour of the solutions
to Eqs.~\ref{eqn:ODE1} and \ref{eqn:ODE2}, for a given value of the
parameter $A$. The arrows on each curve show the direction of time
evolution from the initial state. For prograde orbits we 
restrict ourselves to studying the region,
 $0 \leq \tilde n \leq 10, 0 \leq \tilde \Omega \leq
10$, in the $(\tilde n,\tilde \Omega)$--plane. This is because
$\Omega_{1} = \sqrt{Gm_{12}/R_{1}^{3}}$ corresponds to stellar breakup
velocity and $n \geq \sqrt{Gm_{12}/R_{1}^{3}}$ means that the planet would be orbiting at,
or beneath, the stellar surface. For a HJ with a mass of $M_{J}$ orbiting a star
of mass $M_{\odot}$, $C\sim 0.01$, so $\tilde n \simeq 10$
corresponds to an orbital semi-major axis of $a \simeq 0.01$
AU, and $\tilde n \simeq 0.1$ corresponds to $a \simeq 0.2$
AU, so these plots represent the full range of orbits of the HJs. 
The top plot in Fig.~\ref{fig:withMBplots} is for $A = 100$, which
corresponds to canonical magnetic braking for a G/K star ($\gamma=1$)
and $Q_{1}^{\prime}$ of $10^{6}$.

In the absence of magnetic braking ($A=0$) we recover the standard
tidal evolution equations for a coplanar, circular orbit. These have
been well studied in the literature (e.g.\ \citealt{Counselman1973};
\citealt{Hut1981}). These equations have an equilibrium of
coplanarity and corotation i.e.\ $i = 0$ and $\tilde \Omega=\tilde n$,
where the orbital inclination (or stellar obliquity) $i$ is
defined by $\cos i = \hat{\mathbf{\Omega}}_{1} \cdot \hat{\mathbf{h}}$
and $i \geq 0$. The system
will approach this equilibrium if both the spin angular momentum 
is less than a quarter of the total angular
momentum, and the total angular momentum 
exceeds some critical value (\citealt{Hut1980}; \citealt{Greenberg1974}).
With no braking, orbits initially outside 
corotation ($\tilde \Omega>\tilde n$) are not subject to tidally induced orbital decay, and
asymptotically approach a stable equilibrium $\tilde \Omega = \tilde
n$ for $\tilde n \leq 3^{-\frac{3}{4}}$. Orbits initially inside
corotation can evolve in two different ways,
depending on the stability of the equilibrium state on the solution's closest
approach to $\tilde \Omega = \tilde n$. If $\frac{d\tilde \Omega}{d\tilde t} >
\frac{d\tilde n}{d\tilde t} > 0$ near corotation, then $\tilde n \leq 3^{-\frac{3}{4}}$, and the equilibrium
state is locally stable (though no such curves are plotted in
Fig.~\ref{fig:withMBplots}, since they occur only in the far bottom
left of the plot, near the origin). This is when the corotation radius moves inwards faster than the orbit shrinks
due to tidal friction, which can result in a final stable equilibrium
state for the system if the corotation radius ``catches up'' with the
planet. On the other hand, orbits inside corotation for which this condition
is not satisfied are subject to tidally induced orbital decay, since tidally
induced angular momentum exchange enhances the difference between
$\tilde \Omega$ and $\tilde n$, which leads to further orbital evolution, and the
spiralling in of the planet. This evolution can be seen from the
dot--dashed lines in the bottom plot in Fig.~\ref{fig:withMBplots}.

Including magnetic braking ($A \neq 0$) means that $\tilde \Omega = \tilde
n$ is no longer an equilibrium state, and the total angular
momentum of the system is not conserved. For an orbit initially not subject
to spiralling into the star via tidal transfer of angular momentum
from orbit to spin ($\tilde \Omega \geq \tilde n$) we see from the top
of Fig.~\ref{fig:withMBplots}, that magnetic
braking will spin the star down so that the planet finds itself inside
the corotation radius of the star. Passing through corotation changes the sign of the tidal
torque and causes the planet to spiral into the star. Note
that, if we ignore the age of the system, any bound orbit will eventually decay
in a finite time since the system has no stable equilibrium. The
effect of magnetic braking is to increase the minimum semi-major axis at which the orbit is
not subject tidally induced orbital decay over the nuclear lifetime of the star.

This means that an initially rapidly rotating G--type star hosting a close--in Jupiter
mass companion will lose significant spin angular momentum through magnetic braking (over a
time $\sim \tau_{\mathrm{mb}}$). During this stage of spin-down the spin frequency of the star
may temporarily equal the orbital frequency of its close--in planet, but the rate of angular momentum
loss through magnetic braking will exceed the tidal rate of transfer of
angular momentum from orbit to spin. The stellar spin
continues to drop well below synchronism until the efficiency of
transfer of tidal angular momentum from orbit to spin can compensate
or overcompensate for the braking. If $\frac{d\tilde \Omega}{d\tilde t} >
\frac{d\tilde n}{d\tilde t} > 0$ inside corotation, then tides will
act to spin up the star, though the timescale for this to cause significant spin--up may be
much longer than the stellar lifetime, and this only occurs if the orbit has
sufficient angular momentum to noticeably spin up the star. Otherwise,
the planet continues to spiral inwards once it moves inside
corotation, and $\tilde \Omega
\simeq$ const.

\begin{figure}
  \begin{center}
    \subfigure{\includegraphics[width =
	0.495\textwidth]{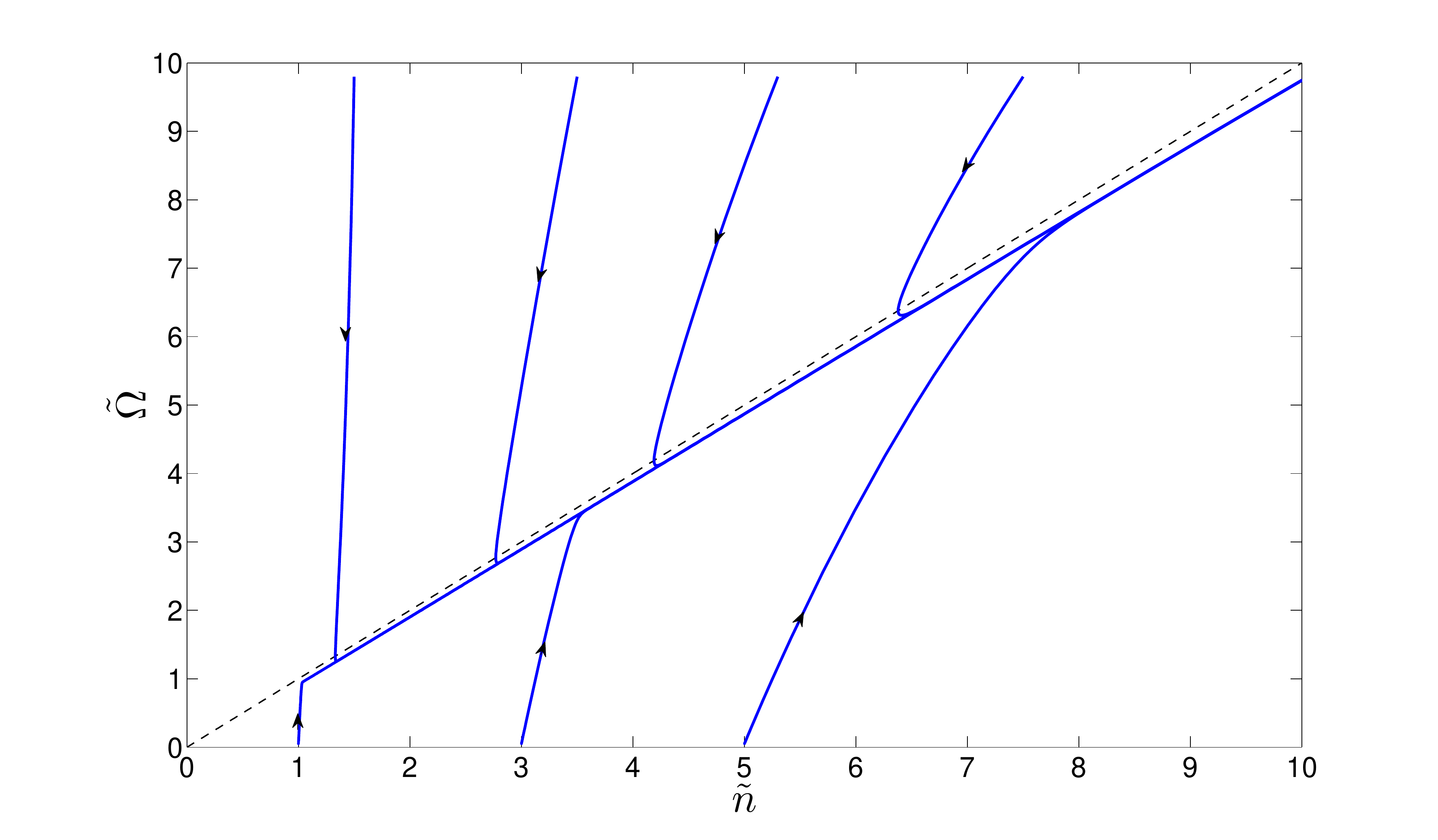}}
  \end{center}
  \caption{($\tilde n,\tilde \Omega$)--plane showing the effects of reducing the stellar mass fraction
    participating in angular momentum exchange with the orbit while the
    braking rate is unchanged, as has
    been proposed for an F star like $\tau$ Boo. $A = 100$, $\epsilon_{\star} = 10^{-2}$}
  \label{FigmbFstar}
\end{figure}

So far we have considered the whole star to participate in tidal angular momentum
exchange with the orbit. For an F-dwarf (like
$\tau$ Boo), it has been proposed that only the outer convective
envelope (of mass fraction $\sim \epsilon_{\star}$) participates
in angular momentum exchange with the orbit 
(\citealt{Marcy1997}; \citealt{DobbsLin2004}; \citealt{Donati2008}). If the core and envelope
of such a star can decouple, then tides would only have to spin up the outer
layers of the star, which would reduce the spin--up time by $\sim \epsilon_{\star}$. In this case
the system could remain in a state with $\tilde \Omega \simeq \tilde
n$ just inside corotation, with the resulting torque on the orbit
small. This may explain the spin--orbit synchronism of stars such as
$\tau$ Boo, as noted by \cite{DobbsLin2004}. Fig.~\ref{FigmbFstar} shows the phase
plane for a simplified system in which the moment of inertia of the
star acted on by tides is reduced by a factor $\epsilon_{\star} \sim 10^{-2}$, but the
braking rate is unchanged i.e. we multiply the first term on the
right-hand side of
Eq.~\ref{eqn:ODE1} by $\epsilon^{-1}_{\star}$. Note that this may be too
simple a model to describe such core--envelope decoupling, and we have
ignored associated changes to the braking rate.

\subsection{Extending the analysis to inclined circular orbits}

We can extend the simplified system of equations analysed in the previous
section to arbitrary inclination ($i$) of the orbital plane
with respect to the equatorial plane of star:

\begin{eqnarray}
\label{eqn:ODEinc1}
\frac{d\tilde \Omega}{d\tilde t} &=&
\tilde n^{4}\left[\cos i
-\frac{\tilde \Omega}{2\tilde n}\left(1+\cos^{2} i \right)\right] -
A \: \tilde \Omega^{3}, \\
\label{eqn:ODEinc2}
\frac{d\tilde n}{d\tilde t} &=& 3 \: \tilde
  n^{\frac{16}{3}}\left[1-\frac{\tilde \Omega}{\tilde n}\cos i
  \right], \\
\frac{d i}{d\tilde t} &=& -\tilde n^{4}\tilde \Omega^{-1}\sin i \left[1-
  \frac{\tilde \Omega}{2\tilde n}\left(\cos i - \tilde n^{\frac{1}{3}}\tilde \Omega
  \right)\right],
\label{eqn:ODEinc3}
\end{eqnarray}
For small inclination, Eq.~\ref{eqn:ODEinc3} reproduces
Eq.~13 from \cite{Hut1981}, with the exception that we have used a constant
$Q_{1}^{\prime}$ rather than a constant time--lag in the equations
(i.e. replace time lag $\tau$ by $\frac{1}{2}\frac{3}{2k_{1}nQ^{\prime}_{1}}$, and
note that $k_{1}$ is twice the apsidal motion constant of the star).

From Eq.~\ref{eqn:ODEinc2} the orbit begins to decay if \begin{eqnarray} \tilde
  \Omega \cos i < \tilde n, \end{eqnarray} which is always satisfied for a retrograde orbit
($i\geq 90^{\circ}$). This is just a generalisation of the corotation
condition $\tilde \Omega =\tilde  n$ to a non--coplanar orbit. The
  inclination grows
if \begin{eqnarray} \tilde \Omega > \tilde \Omega_{crit} =
  2\tilde n \left(\cos i -\tilde n^{\frac{1}{3}}\tilde \Omega
  \right)^{-1}, \label{incexc}
\end{eqnarray} where we have assumed the quantity in brackets is
positive i.e. $\cos i >\tilde n^{\frac{1}{3}}\tilde \Omega$. 
This agrees with the condition from \cite{Hut1981}
when $i\sim 0$. 
For sufficiently close--in orbits that tidal friction
is important, magnetic braking will rapidly spin down the star such
that this condition is not satisfied, so we can safely conclude that the
inclination is not likely to grow appreciably by tidal friction. When
this condition is not satisfied, the inclination decays to zero, on a
timescale $\tau_{i}$ (see next section). Note that $i=180^{\circ}$ is an
unstable equilibrium value of the inclination.

\begin{figure}
  \centering
  \subfigure{\label{Figmbi1}\includegraphics[width=0.495\textwidth]{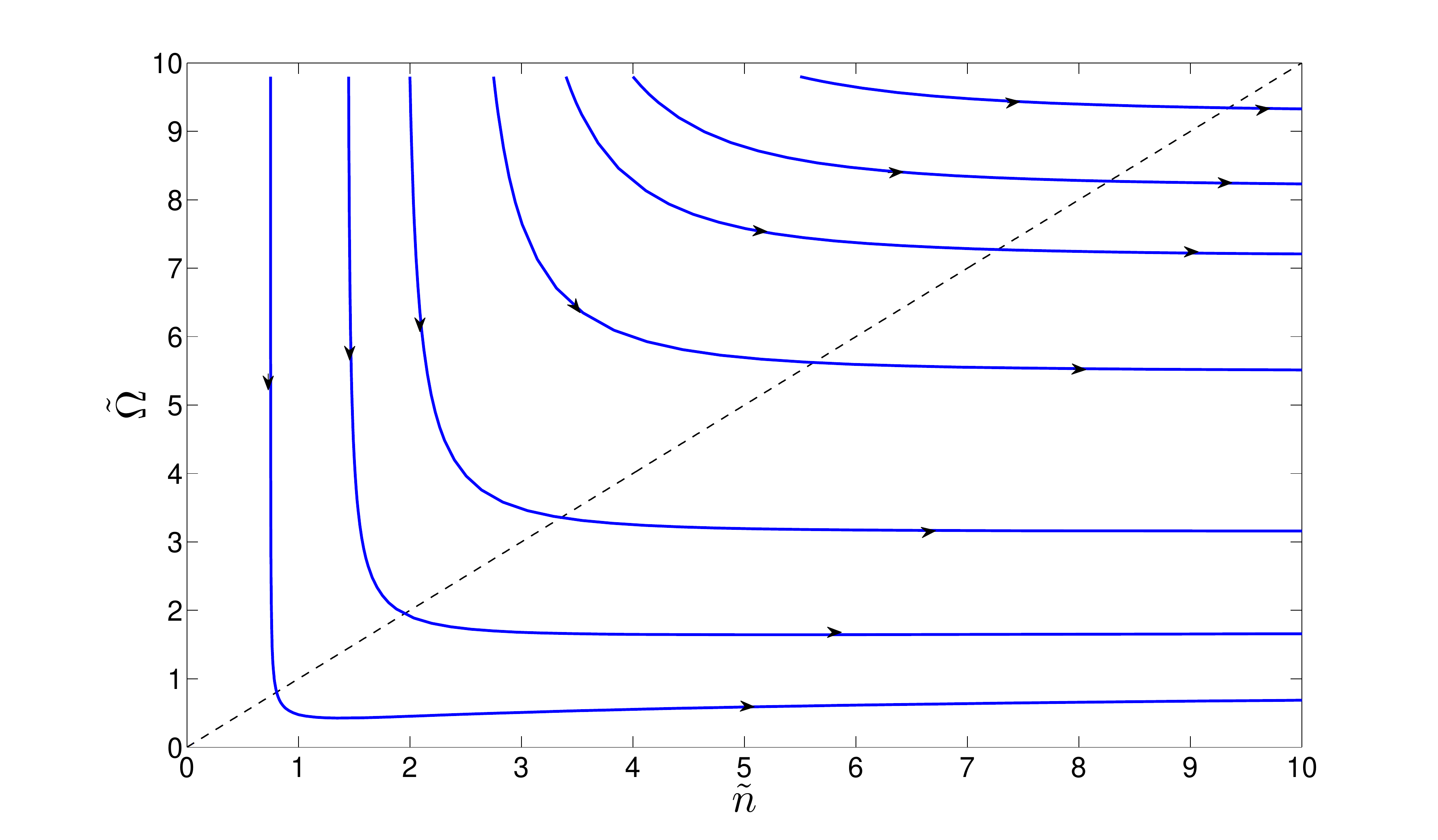}}\\
  \subfigure{\label{Figmbi2}\includegraphics[width=0.495\textwidth]{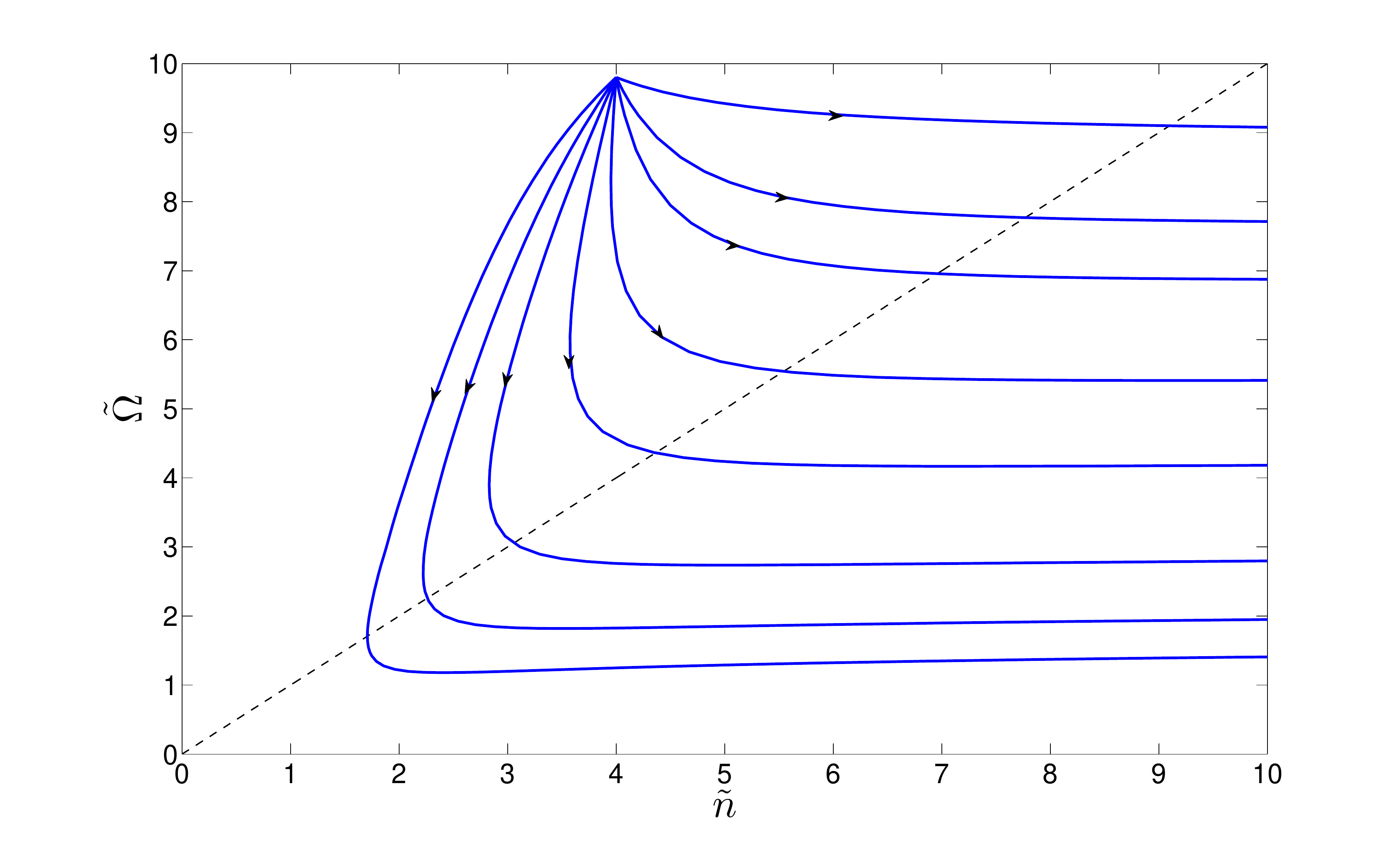}}
  \caption{Top: ($\tilde n,\tilde \Omega$)--plane for an orbit with an
    initial $i=90^{\circ}$, with $A = 10$, where the dashed line corresponds to
    $\tilde \Omega=\tilde n$. This is similar to Fig.~\ref{Figmb1} for a circular orbit,
    except that the orbit decays once $\tilde \Omega
    \cos i < \tilde n$, which can occur above the dashed line in contrast with
    the circular case. The evolution of $i$ is not plotted, and is in general different
    for each curve, but is found to decay once the stellar spin drops below that given by
    Eq.~\ref{incexc}.
    Bottom: various initial inclinations, with $A = 10$, where the dashed line corresponds to
    $\tilde \Omega=\tilde n$. The bottom trajectory has $i=0$, and
    the inclination is increased in steps towards the top curve, which
    has $i=180^{\circ}$, to
    illustrate the behaviour.
    Note that orbits with larger initial $i$ decay for smaller
    $\tilde n$, once $\tilde \Omega \cos i < \tilde n$ is satisfied.}
  \label{Figmbi}
\end{figure}

In the absence of magnetic braking we recover the evolution considered
  by \cite{Greenberg1974},  so we will now concentrate on the inclusion
  of magnetic braking. The top figure in Fig.~\ref{Figmbi} plots the ($\tilde n,\tilde \Omega$)--plane
for $A = 10$ for an initial inclination of $i=90^{\circ}$. A smaller value
of $A$ is chosen over the previous section in order to show the
effects of tides more clearly, since reducing $A$ is equivalent to
reducing $Q^{\prime}$. The evolution of $i$ is not plotted 
(and will in general be different
for each curve), but is found to
decay once the stellar spin decays sufficiently that Eq.~\ref{incexc}
is not satisfied. The bottom figure in Fig.~\ref{Figmbi} shows the effect of
increasing the inclination in steps to illustrate the behaviour as $i$
is increased, on various curves with otherwise the same
initial conditions, with $A = 10$. The orbit begins to decay
for smaller $\tilde n$, and decays at a faster rate as $i$ is
increased. This peaks for a perfectly retrograde
orbit ($i=180^{\circ}$), with anti--parallel spin and orbit, where the rates of change of
spin and orbital angular frequencies are maximum. Fig.~\ref{Figmbi}
shows that the orbit generally begins to decay outside corotation, 
once $\tilde \Omega \cos i < \tilde n$. This has implications for the
tidal evolution of close--in planets on inclined orbits, in that if
this condition is satisfied, the planet will be undergoing tidally
induced orbital decay -- though the inspiral time may be longer than
the expected stellar lifetime.

In this section we have seen that magnetic braking can only be reasonably
neglected for a coplanar orbit when $\Omega \ll n$. For an inclined
orbit, this condition must be generalised to $\Omega \cos i \ll n$ -
obvious from Eq.~\ref{eqn:ODEinc2}. Neglecting $\dot {\bmath\omega}_{\mathrm{mb}}$ for stars
for which this condition is not satisfied can result in a
qualitatively different evolution, as already seen in
Fig.~\ref{Figmb2} for a coplanar orbit.

\section{Tidal evolution timescales}
\label{timescales}

It is common practice to interpret the effects of tidal evolution in terms of simple
timescale estimates. The idea behind these is that if the rate of change of a
quantity $X$ is exponential, then $\dot X/X$ will be a constant, so we
can define a timescale $\tau_{X} = X/\dot X$. If $\dot X/X \ne$ 
const, then these may not accurately represent the
evolution. Here we reproduce the timescales that can be derived from the equations in Appendix~\ref{Appendix}.

A tidal inspiral time can be calculated from the equation for $\dot
a$, by considering only the effects of the tide raised on the star by
the planet (not
unreasonable since $I_{2} \ll I_{1} \sim m_{2}a^{2}$). Here $\dot a/a
\sim a^{-13/2} \ne $ const, so a more accurate
estimate of the inspiral time for a circular coplanar orbit is
\small{
\begin{eqnarray*} \hspace{-10.0pt} \tau_{a} \hspace{-9.0pt} &\equiv& \hspace{-10.0pt} -\frac{2}{13}\frac{a}{\dot a} \\
  &\simeq& \hspace{-10.0pt} 12.0 \;\mbox{Myr}
\left(\frac{Q_{1}^{\prime}}{10^{6}}\right)\left(\frac{m_{1}}{M_{\odot}}\right)^{\frac{8}{3}}
\left(\frac{M_{J}}{m_{2}}\right)
\left(\frac{R_{\odot}}{R_{1}}\right)^{5}\left(\frac{P}{1\mathrm{d}}\right)^{\frac{13}{3}}\left(1-\frac{P}{P_{\star}}\right)^{-1}
\end{eqnarray*} } \normalsize
Here $P$ and $P_{\star}$ are the orbital and stellar spin periods, respectively.
We have already seen in \S \ref{phaseplaneanalysis} that it is unreasonable to assume that
$\Omega$ is fixed unless $\Omega \ll n$, due to magnetic braking.

If the orbit is inside corotation, angular momentum will be
transferred from the orbit to the spin of the star, giving a tidal spin--up time of
\begin{eqnarray*} \tau_{\Omega_{1}} &\equiv& -\frac{\Omega_{1}}{\dot
    \Omega_{1}} \simeq \frac{13\tau_{a}}{2\;\alpha},
\end{eqnarray*} 
where $\alpha=\frac{\mu h}{I \Omega}$ is the ratio of orbital to spin angular momentum. For the
HJ problem, $\tau_{\Omega_{1}}\geq \tau_{a}$ since $\alpha\sim O(1)$, which neglects the spin--down effects of magnetic braking. The
planetary spin $\Omega_{2}$ will tend to synchronise much faster,
since the moment of inertia of the planet is much less than that of
the orbit (by $\sim 10^{5}$), and will not be considered further
i.e.\ we assume $\Omega_{2} = n$.

A circularisation time can be derived from the equation for
$\mathbf{\dot e}$, and is given for a coplanar orbit by
\small
\begin{eqnarray*}
  \tau_{e} &\equiv& -\frac{e}{\dot e} \\
   &\simeq& 16.8 \;\mbox{Myr} \left(\frac{Q_{1}^{\prime}}{10^{6}}\right)\left(
  \frac{m_{1}}{M_{\odot}}\right)^{\frac{8}{3}}\left(\frac{M_{J}}{m_{2}}\right)
  \left(\frac{R_{\odot}}{R_{1}}\right)^{5}\left(\frac{P}{1\mathrm{d}}\right)^{
    \frac{13}{3}} \\ && \hspace{-10.0pt}
   \times\left[\left(f_{1}(e^{2})-\frac{11}{18}
     \frac{P}{P_{\star}}f_{2}(e^{2})\right)+
    \beta \left(f_{1}(e^{2})-\frac{11}{18}f_{2}(e^{2})\right)\right]^{-1}
\end{eqnarray*}
\normalsize
where we have included both the stellar and planetary tides, as these
have been shown to both contribute to the tidal evolution of $e$
\citep{JacksonI2008}.
Note that $\dot e/e \sim $ const only if $P\sim$ const and $e \ll 1$, where $f_{1,2}(e^{2})\simeq
1$. The factor 
\begin{eqnarray*} \beta =
\frac{Q_{2}^{\prime}}{Q_{1}^{\prime}}\left(\frac{m_{1}}{m_{2}}\right)^{2}\left(\frac{R_{2}}{R_{1}}\right)^{5}
\sim 10 \frac{Q_{2}^{\prime}}{Q_{1}^{\prime}}
\end{eqnarray*} for the HJ problem.

If the orbital and stellar equatorial planes are misaligned, then
dissipation of the tide raised on the star by the planet would align them on a timescale
\small
\begin{eqnarray*}
  \tau_{i} &\equiv& -\frac{i}{\frac{di}{dt}} \\
  &\simeq& 70 \; \mbox{Myr} \: 
  \left(\frac{Q_{1}^{\prime}}{10^{6}}\right)
  \left(\frac{m_{1}}{M_{\odot}}\right)
  \left(\frac{M_{J}}{m_{2}}\right)^{2}
  \left(\frac{R_{\odot}}{R_{1}}\right)^{3}\left(\frac{P}{1\mathrm{d}}\right)^{4} \\
&& \times\left(\frac{\Omega_{1}}{\Omega_{0}}\right)
  \left[1-
    \frac{P}{2P_{\star}}\left(1 - \frac{1}{\alpha}
    \right)\right]^{-1}
\end{eqnarray*} \normalsize
where we have assumed that the orbit is circular and made the
small $i$ approximation. We
take $\Omega_{0} = 5.8 \times 10^{-6}\mbox{s}^{-1}$, which corresponds
to a spin period of $\sim 12.5$ d.

The validity of these timescales to accurately represent the tidal
evolution of the orbital and rotational elements is an important
subject of study, since these timescales are commonly applied to
observed systems. In a recent paper, \cite{JacksonI2008} found that it
is essential to consider the coupled evolution of $e$ and $a$ in order
to accurately model the tidal evolution, and that both the stellar and
planetary tides must be considered. They showed that the actual change
of $e$ over time can be quite different from simple
circularisation timescale considerations, due to the coupled evolution
of $a$. In the following we will consider the validity of the
spin--orbit alignment timescale to accurately model tidal evolution of $i$.

\section{Numerical integrations of the full equations for an inclined orbit}

We perform direct numerical integrations of the equations in Appendix \ref{Appendix} with a 4th/5th order
Runge-Kutta scheme with adaptive stepsize control, using a scheme with
Cash-Karp coefficients similar to that described in
\cite{Press1992}. Our principal aim is to study inclination evolution and to
determine the accuracy of the spin--orbit alignment timescale $\tau_{i}$ for
close--in planets.

We choose a ``standard'' system of a HJ in orbit around an FGK star. We
have $m_{1} = M_{\odot}, m_{2} = M_{J}, R_{1} = R_{\odot}, R_{2} =
R_{J}$, and we modify\footnote{We take $\mathbf{\Omega}_{1}
    \cdot \mathbf{e} = 0$, though this need not be assumed. We find
  negligible difference between integrations for which $\mathbf{\Omega}_{1}
    \cdot \mathbf{e} = 0$ and $\mathbf{\Omega}_{1}
    \cdot \mathbf{e} \ne 0$} $a$, $i$, $e$. We choose the initial ratio
$\Omega_{1}/n=10$, and include magnetic braking in all simulations,
setting $\gamma = 1$ unless stated otherwise. We choose
$Q_{1}^{\prime}=Q_{2}^{\prime}=10^{6}$, and $\Omega_{2}/n=1$ unless stated otherwise.
The dimensionless radii of gyration are chosen to be $r^{2}_{g1} = 0.076$ and
$r^{2}_{g2} = 0.261$, which are values appropriate for polytropic
stellar and planetary models with
respective indices 3 and 1.

\subsection{Inclusion of magnetic braking and the importance of
  coupled evolution of $a$ and $i$}

\begin{figure*}
  \begin{center}
    \subfigure[$a$ evolution for $a=0.05$ AU]{\label{Coupleda0.05-a}
      \includegraphics[width=0.485\textwidth]{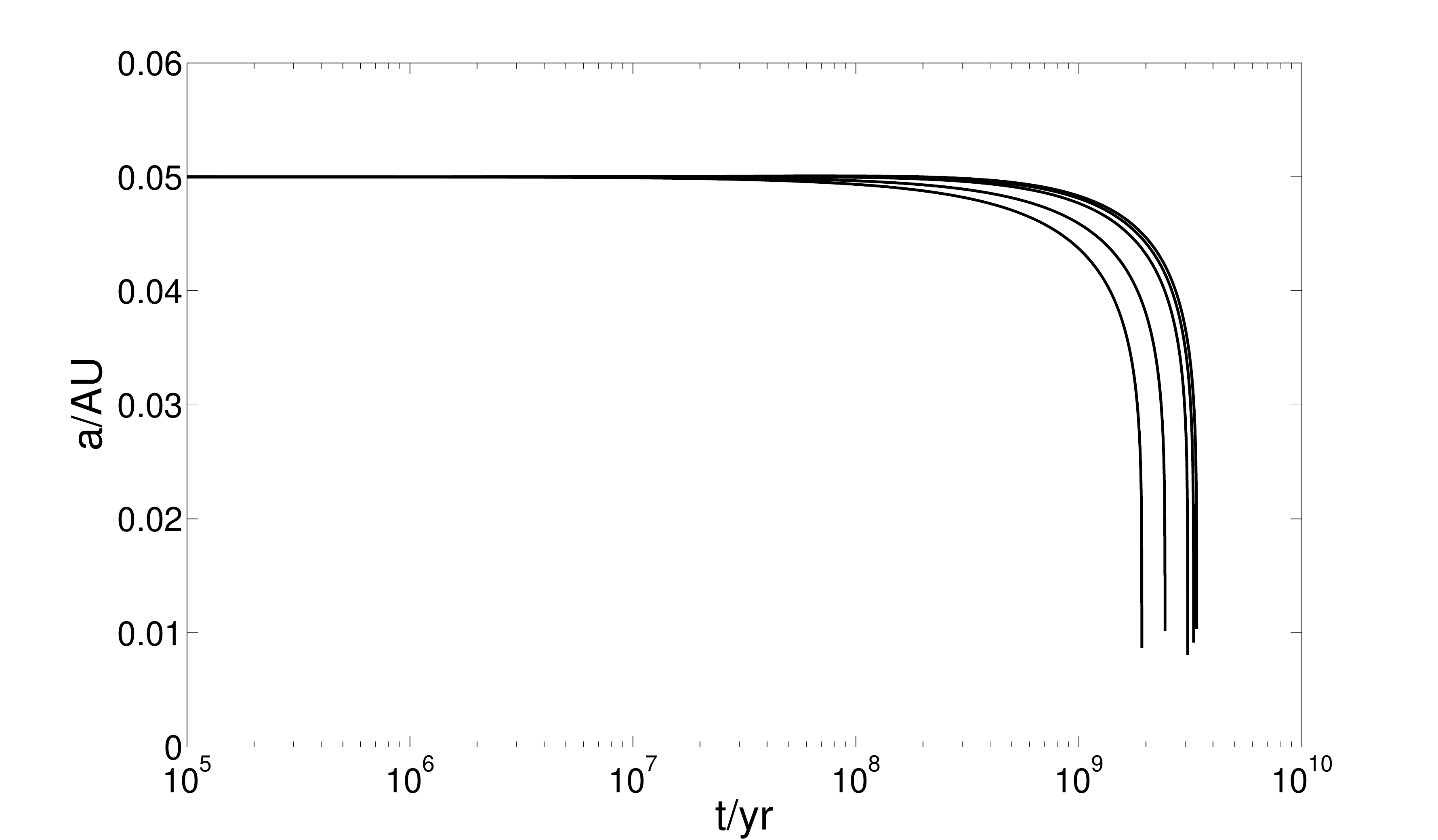}}
    \subfigure[$\cos i$ evolution for $a=0.05$ AU]{\label{Coupleda0.05-i}
      \includegraphics[width=0.485\textwidth]{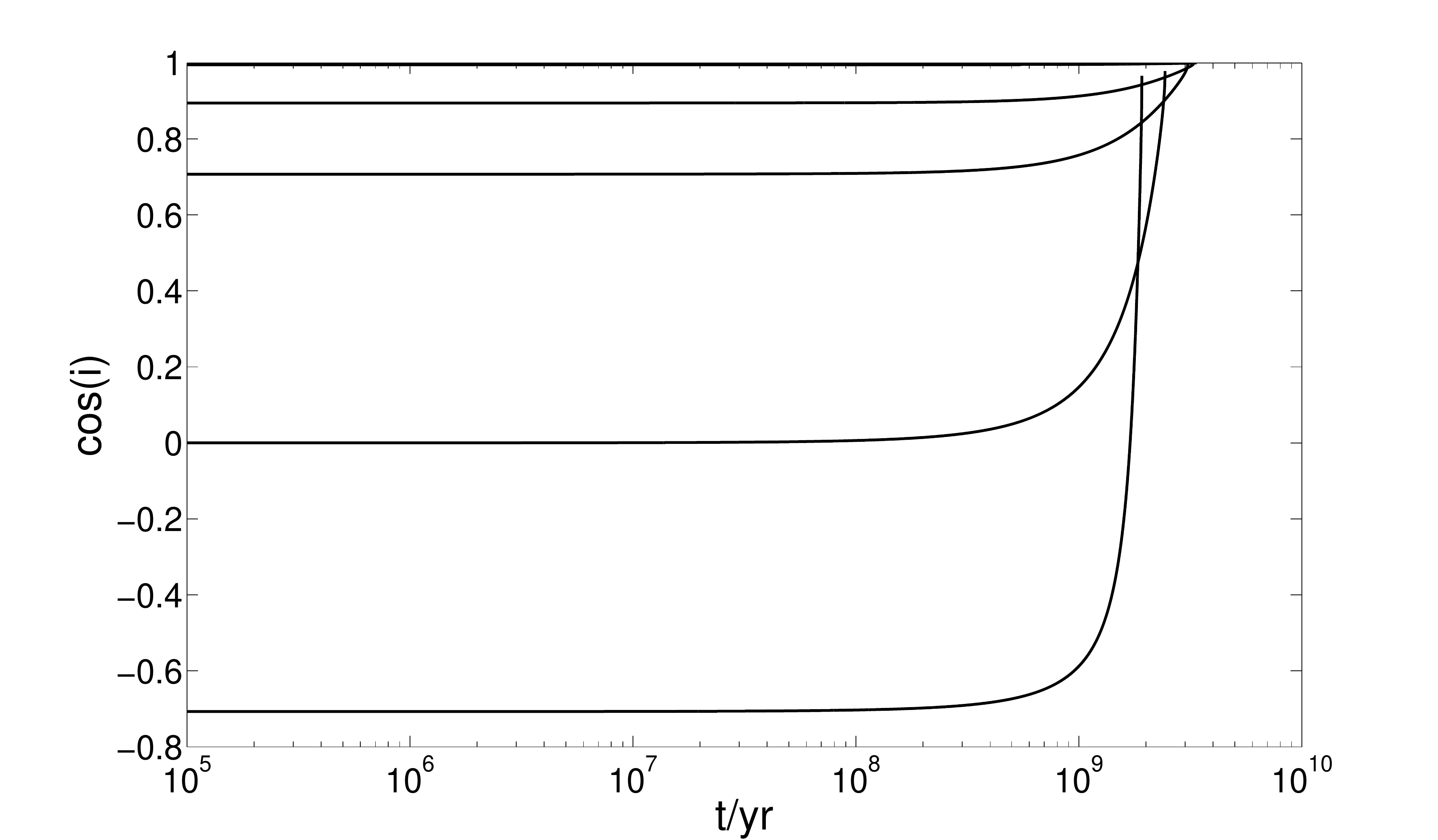}}
    \subfigure[$a$ evolution $a=0.08$ AU]{\label{Coupleda0.08-a}
      \includegraphics[width=0.485\textwidth]{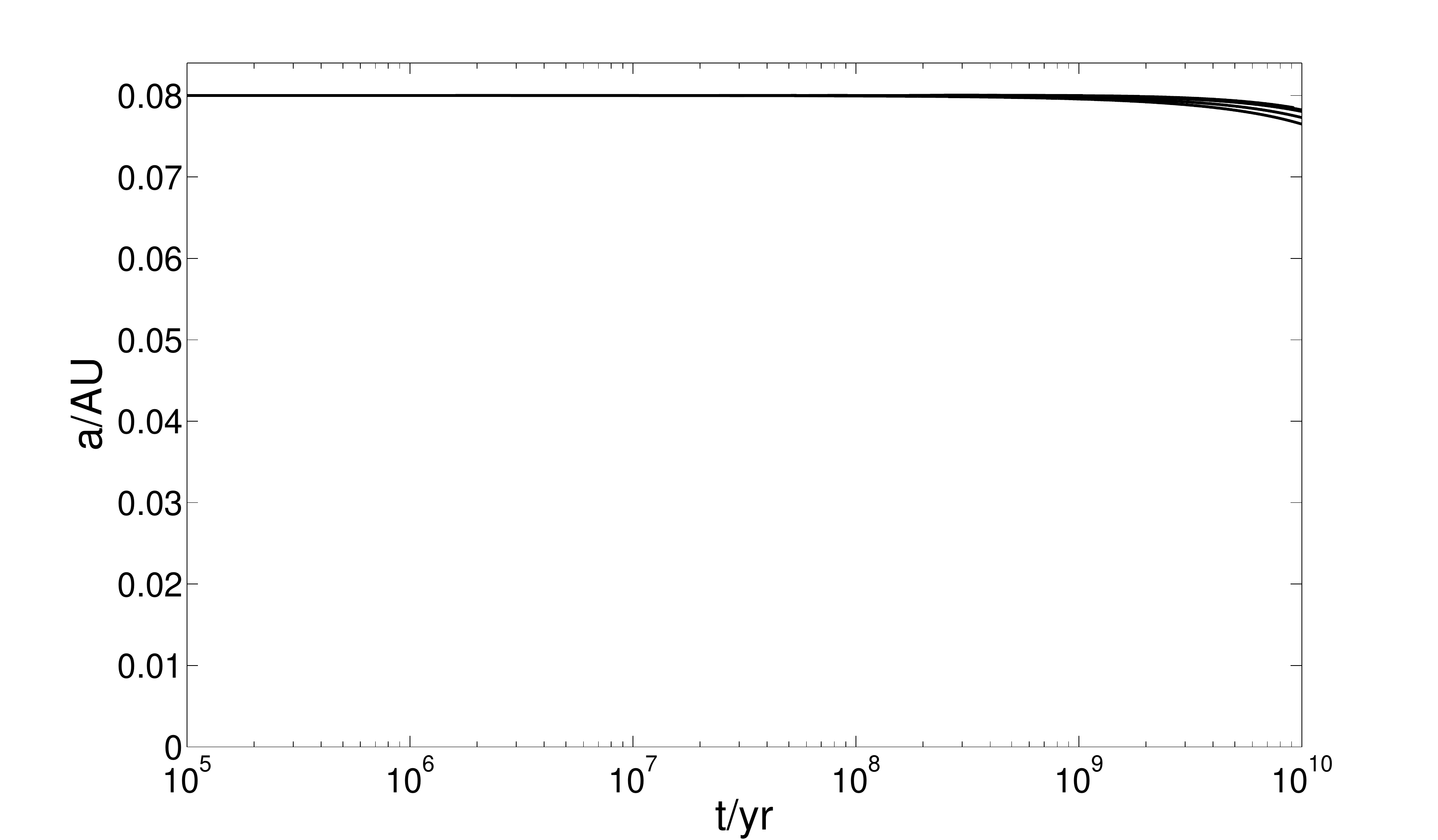}}
    \subfigure[$\cos i$ evolution $a=0.08$ AU]{\label{Coupleda0.08-i}
      \includegraphics[width=0.485\textwidth]{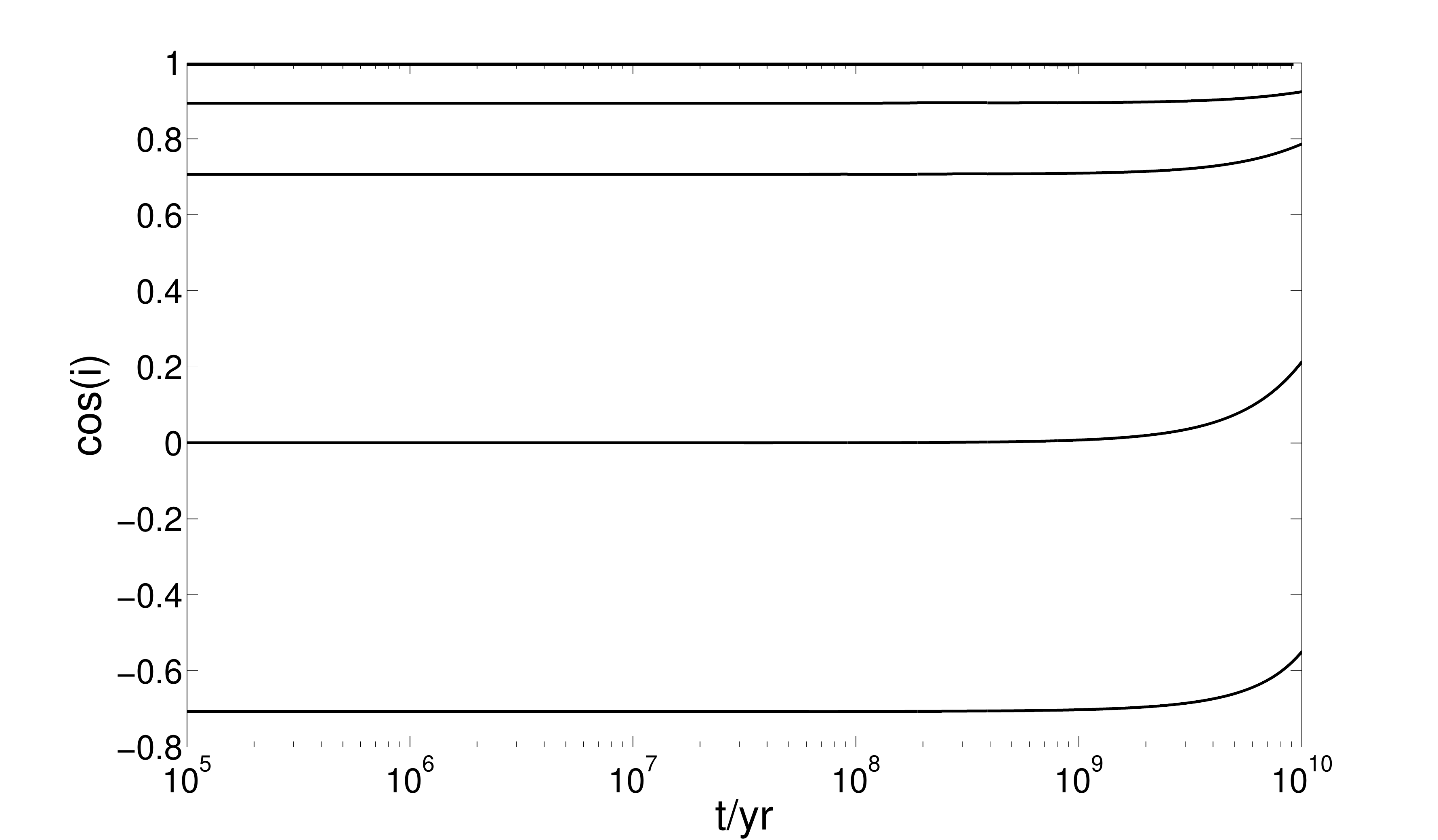}}
  \end{center}
  \caption{Tidal evolution for a circular, inclined orbit at $a=0.05$ AU and
    $a=0.08$ AU, with various initial inclinations:
    $i=6^{\circ},26^{\circ},45^{\circ},90^{\circ},180^{\circ}$. 
    (a) and (c) show their respective semi-major axis
    evolutions (with the highest inclination orbit decaying first --
    the bottom curve -- and the lowest inclination orbit decaying last
    -- the top curve),
    and (b) and (d) show the respective inclination evolution for these
    systems. The outer orbit $a$ changes only slightly over $10$ Gyr, and $i$ evolves as
    expected from simple estimates of $\tau_{i}$. The inner orbit, on the
    other hand, is tidally shrunk, and this reduction in $a$ means
    that the true evolution is much faster than the simple timescale
    estimates predict. This highlights the importance of
    considering coupled evolution of $a$ and $i$.}
  \label{Fig:Coupleda-i}
\end{figure*}

For a prograde orbit ($i < 90^{\circ}$) initially outside corotation,
$\dot {\bmath{\omega}}_{\mathrm{mb}}$ rapidly spins the star down
sufficiently to ensure that inclination is not excited through tidal
friction (so that Eq.~\ref{incexc} is not satisfied), and the
inclination begins to
decay. Subsequent spin--down moves the
orbit--projected corotation radius beyond the orbit of the planet (so
that $\Omega \cos i < n$), and the resulting tidal
inspiral accelerates as the difference between $\Omega \cos i$ and $n$ is
enhanced. The associated reduction in $a$ increases the rate of
stellar spin--orbit alignment. Thus the inclusion of $\dot
{\bmath{\omega}}_{\mathrm{mb}}$ increases the
rate of alignment, and reduces $\tau_{i}$ from the simple
estimate, which ignores magnetic braking and coupled $a$ and $i$ evolution. 

The effect of $\dot {\bmath{\omega}}_{\mathrm{mb}}$ on a retrograde orbit ($i
\geq 90^{\circ}$) is qualitatively different. A retrograde orbit is always
subject to tidally induced inspiral since $\Omega \cos i < n$ for all $i
\geq 90^{\circ}$. $\dot {\bmath{\omega}}_{\mathrm{mb}}$ acts to reduce $\Omega$, thereby reducing the difference
$|\Omega \cos i - n|$, making the tidal torque smaller. This
acts to \textit{increase} the timescale for alignment of the stellar
spin and orbit, though the effect
is found to be small.

The most important effect of including $\dot {\bmath{\omega}}_{\mathrm{mb}}$
is simply that of reducing the stellar spin sufficiently so that
$\Omega \cos i < n$, where the semi-major axis can then decay through tidal
friction. As $a$ subsequently decreases, the tidal torque increases, resulting in a
faster inclination decay rate. $\dot {\bmath{\omega}}_{\mathrm{mb}}$ can only be
neglected if $|\Omega \cos i| \ll n$. If the orbit is already highly inclined,
such as $i\geq 90^{\circ}$, then the inclusion of magnetic braking
is not so important, since $\Omega \cos i < n$ regardless of the spin rate.

For an orbit initially at $a = 0.05$ AU, the simple estimate of the
stellar spin-orbit alignment timescale gives $\tau_{i} \simeq 2.0
\times 10^{10}$
yrs. Fig.~\ref{Fig:Coupleda-i} shows the evolution of $i$ and $a$ for various initial inclinations for
an orbit at $a = 0.05$ AU and $a=0.08$ AU respectively. The outer
orbit $a$ changes only slightly over $10$ Gyr, and $i$ evolves as
expected from the simple estimate of $\tau_{i}\simeq 3.0 \times
10^{11}$ yrs. The inner orbit, on the
other hand, is subject to tidally induced orbital decay, with an
inspiral time of $1-3$ Gyr, and this reduction in $a$ increases the
rate of inclination evolution. This highlights the importance of
considering coupled evolution of $a$ and $i$, especially for large
initial inclinations, where inspiral occurs for higher stellar spin rates. The
difference between the simple estimate of $\tau_{i}$ and the true timescale can be
up to an order of magnitude different for orbits whose $a$ changes appreciably.

\subsection{Inclined and eccentric orbits}

\begin{figure*}
  \begin{center}
    \subfigure{\label{inceccorba}
      \includegraphics[width=0.485\textwidth]{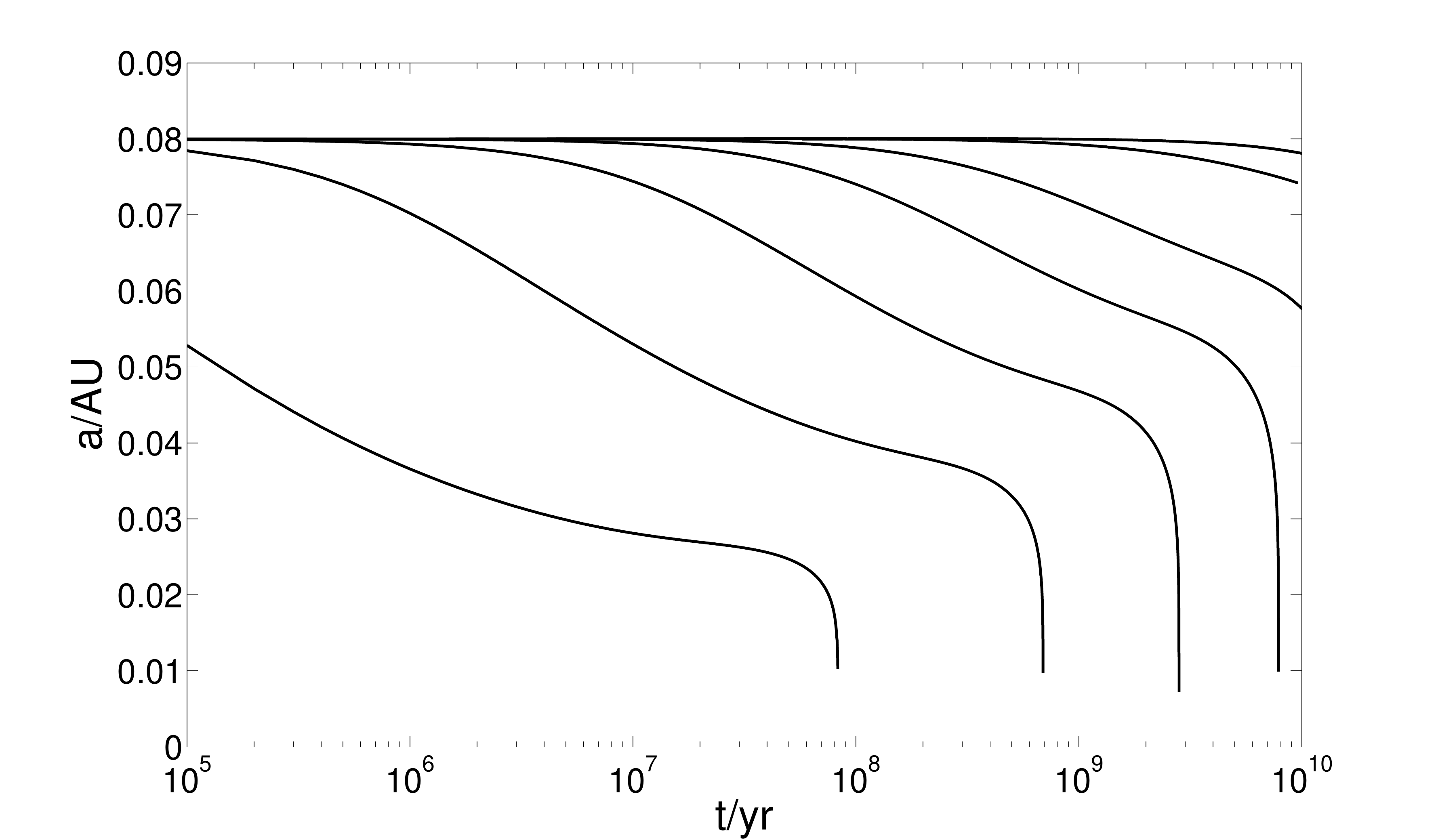}} 
    \subfigure{\label{inceccorbi}
      \includegraphics[width=0.485\textwidth]{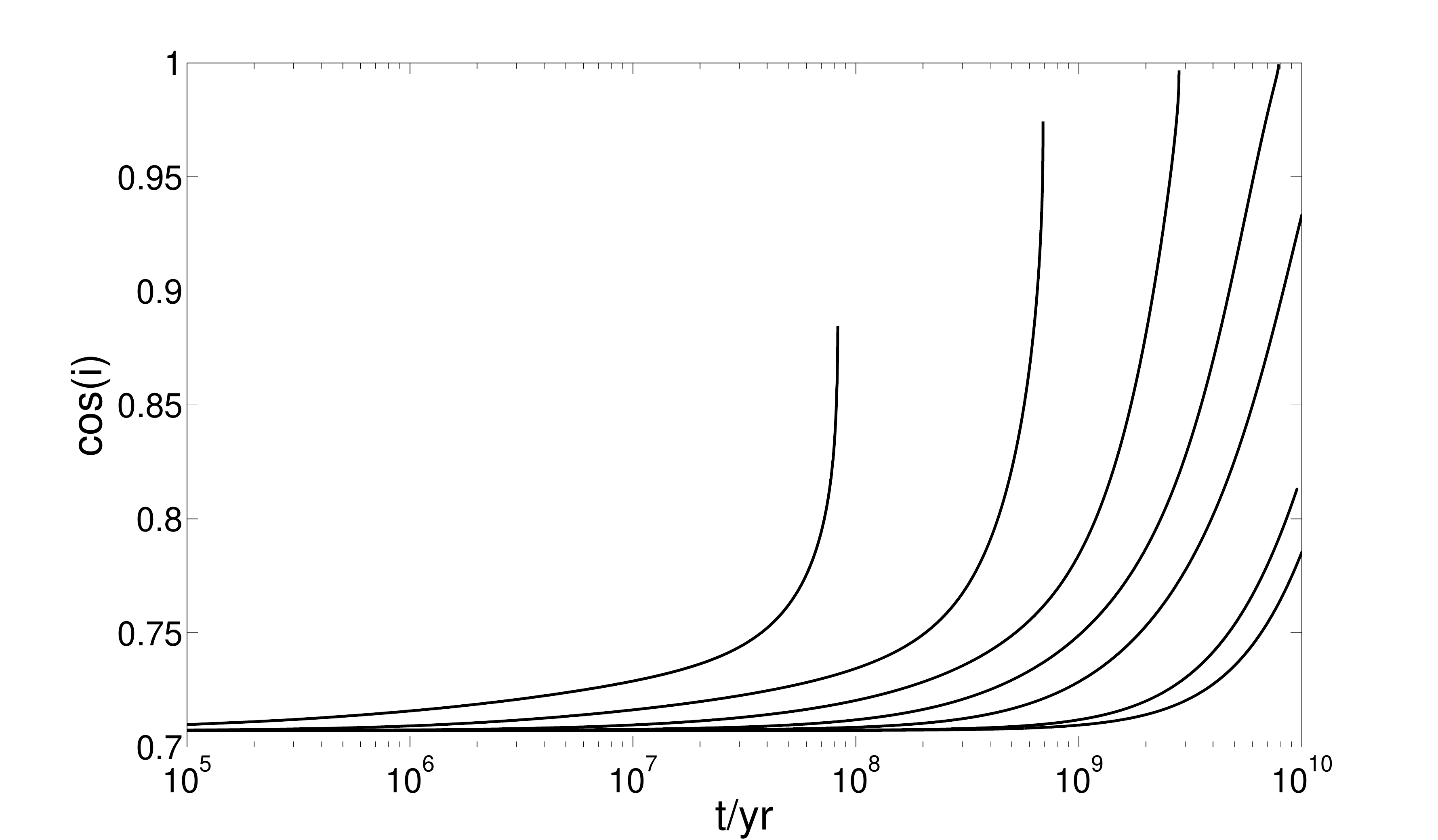}}
    \subfigure{\label{inceccorbe}
      \includegraphics[width=0.485\textwidth]{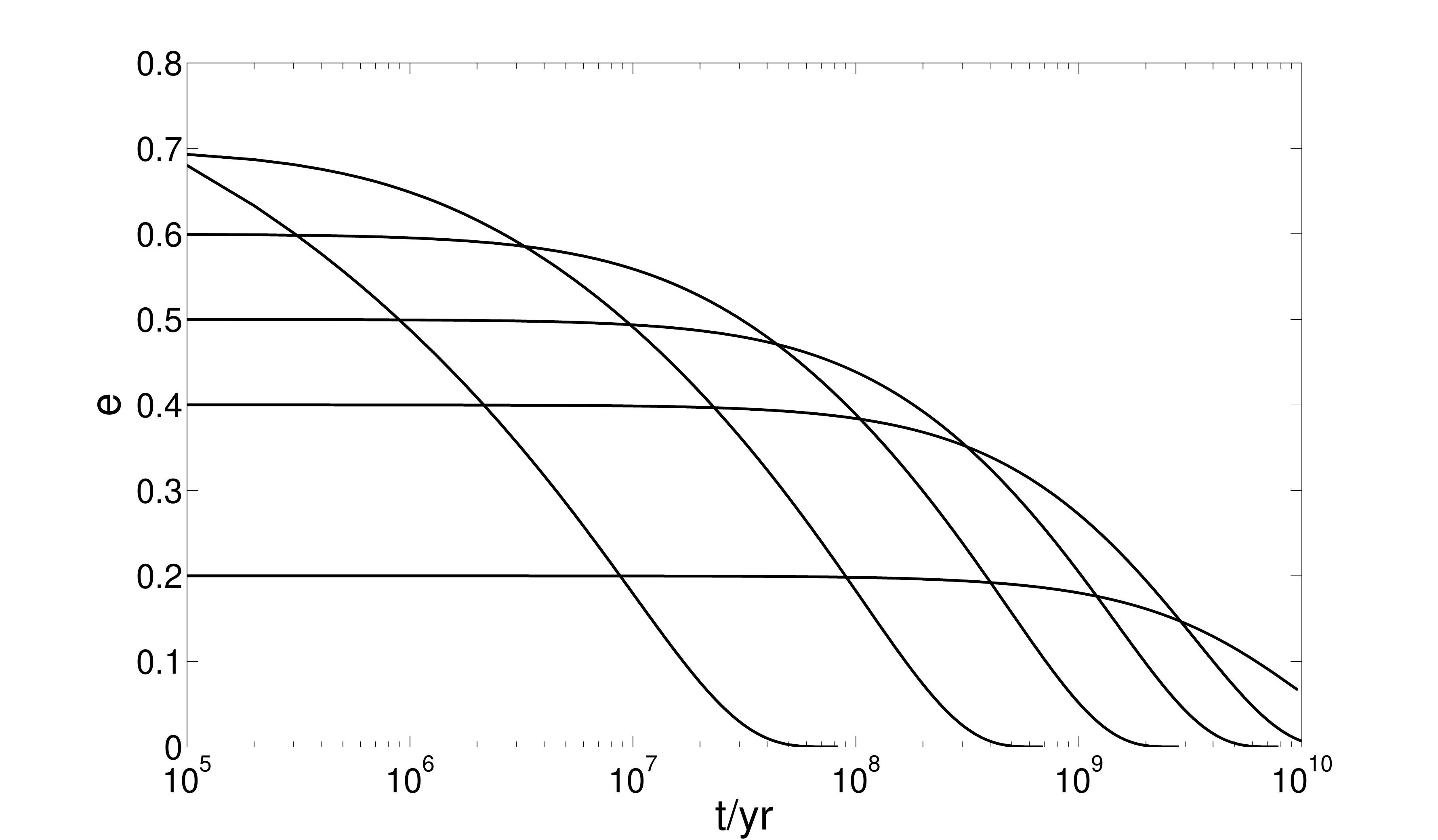}}
  \end{center}
  \caption{Top left: Semi-major axis evolution for an inclined orbit with
    initial $i=45^{\circ}$ at
    $a=0.08$ AU for various initial $e$, with
    $e=0,0.2,0.4,0.5,0.6,0.7,0.8$.
    Top right and bottom: Inclination and eccentricity evolution for the same
    systems. Solutions with the smallest
    initial $e$ have the smallest change in $a$ and $i$. Solutions
    with the largest initial $e$ undergo much more rapid tidal
    evolution (note that for $e=0.8$, $e$ decays to less than $0.7$
    within $10^{5}$ yrs) -- curves can be distinguished by noting that
    the curves corresponding to the fastest evolution have the largest
    initial $e$. Increasing the
    eccentricity can be seen to reduce the inspiral time by up to several orders
    of magnitude over the circular case. In contrast, increasing the
    inclination in Fig.~\ref{Fig:Coupleda-i} only reduces the inspiral
    time by a factor $\sim O(1)$ over a coplanar orbit. Also note that
    $\tau_{e} < \tau_{i}$ for all integrations.}
  \label{Fig:incleccorb}
\end{figure*}

\cite{JacksonI2008} already highlighted the importance of coupled $a$
and $e$ evolution for a coplanar orbit. We will now consider how $e$
might affect $i$ evolution for a non--coplanar orbit, which is the
subject of this section.

A nonzero eccentricity reduces the pericentre distance $r_{p} =
a(1-e)$, which increases the tidal torque over a circular orbit, since
tidal torque $\sim r^{-6}$. Although the planet would spend less time near
pericentre, the torque there is much greater, so dominates
the orbit--averaged torque. We therefore expect
the stellar spin--orbit alignment time to be reduced as we increase $e$. In
addition, we expect that an orbit at large $e$ would more strongly
affect the rate of alignment over a circular orbit, than
one at large $i$ would over a coplanar orbit, because $e$ reduces $r_{p}$, whereas $i$ only changes the
difference $(\Omega\cos i -n)$. The tidal torque $\sim
r^{-6}(\Omega\cos i -n)$, which depends more strongly
on $r$ than $\Omega$. 
This behaviour can be seen in Figs.~\ref{Coupleda0.08-i} and
\ref{inceccorbi} which shows that the ratios of stellar spin--orbit alignment times for an orbit with small $i$ and
large $i$ is $\sim O(1)$, whereas the ratios of stellar spin--orbit alignment times
for an orbit with small $e$ and large $e$ can be up to several orders of
magnitude.

From Fig.~\ref{Fig:incleccorb} we can compare the simple estimate of
$\tau_{i} \simeq 3.0 \times10^{11}$ yrs for an orbit at $a=0.08$ AU, with the
coupled evolution of the orbital and rotational elements from
integration of the full equations. We see that the simple estimate
gives a misleadingly long stellar spin--orbit alignment time compared
with that obtained from integrating the full equations in cases where
$e$ is initially non--negligible -- as a result of
the strong functions of the eccentricity in this model of tidal
friction (see Appendix \ref{Appendix}). This confirms the
conclusion of \cite{JacksonI2008}, in that it is essential to consider
coupled evolution of $a$ and $e$ in order to determine an accurate
system history. We also find that the associated changes in semi-major
axis strongly affect stellar spin--orbit alignment. 
A marginally better estimate for $\tau_{i}$ can be made by replacing
the orbital period with the orbital period around periastron
(equivalent to replacing $a$ by $r_{p}$), though this is still inadequate since it
neglects evolution of $a$.

\subsection{Discussion}

For typical HJs, we find that the stellar spin--orbit alignment
time is comparable to the inspiral time i.e.\ $\tau_{i} \sim
\tau_{a}$. This means that if we observe a planet, then its
survival implies that tides are unlikely to have aligned its
orbit. For planets on an accelerating inspiral
into the star, the rate of inclination evolution will have been much 
lower in the past. Therefore if we observe a planet well inside 
corotation ($\Omega \cos i < n$), with a
roughly coplanar orbit, we can assume that it must have started off
similarly coplanar -- unless we are lucky enough to be observing 
a planet on its final rapid inspiral into the star after it has
undergone most of the evolution, where it is now in a very
short--period orbit, close to being consumed. 

We expect $\tau_{i} \sim \tau_{a}$ when $\alpha \ll 1$ -- which is true
for close--in terrestrial planets -- since $\mathbf{\Omega}$
can be considered fixed, with the inclination changing only due to changes in
$\mathbf{h}$. For typical values of $\alpha \sim O(1)$ for HJs, the inclination
changes due to rotations of both $\mathbf{\Omega}$ and $\mathbf{h}$, 
so the timescales are not exactly the same, but would be expected to
be of the same order of magnitude.

\cite{Hut1981} showed by considering only the tide in the
  star, that the
stellar spin--orbit alignment timescale is longer 
than the circularisation timescale
($\tau_{i} > \tau_{e}$) unless $\alpha > 6$. This was based on
exponential decay estimates for small $e$ and $i$, but nevertheless
holds for the systems integrated in this work, as can be seen from
Fig.~\ref{Fig:incleccorb}. This makes intuitive sense, since
$\alpha \sim O(1)$ means that spin angular momentum is important, and circularisation
involves only a property of the orbit, whereas alignment involves both the spin and
the orbit. For typical HJs $\alpha < 6$; we therefore expect
$\tau_{e} < \tau_{i}$, especially when the eccentricity damping effect
of the tide in the planet
is taken into account, which further enhances
this inequality. The tide in the planet is completely negligible in
changing the stellar obliquity since $I_{2} \ll I_{1} \sim
m_{2}a^{2}$. 

This means that if an orbit is initially inclined and eccentric as a
result of planet--planet scattering or Kozai migration into a short--period orbit, we would
expect the orbit to become circular before
it aligns with the spin of the star. We should therefore observe fewer eccentric orbits than inclined orbits,
\textit{if those systems start with a uniform distribution in
  $e,i$--space}. This also means that if we observe a
close--in planet on a circular orbit with non-zero $i$, we cannot rule out a
non-negligible eccentricity in the past.

\section{Application to observed systems}
\label{Obs}

\subsection{An explanation of the misaligned spin and orbit of XO-3 b}
\label{XO-3}

\begin{figure}
  \begin{center}
    \subfigure{\label{FigXO-3-1}
      \includegraphics[width=0.47\textwidth]{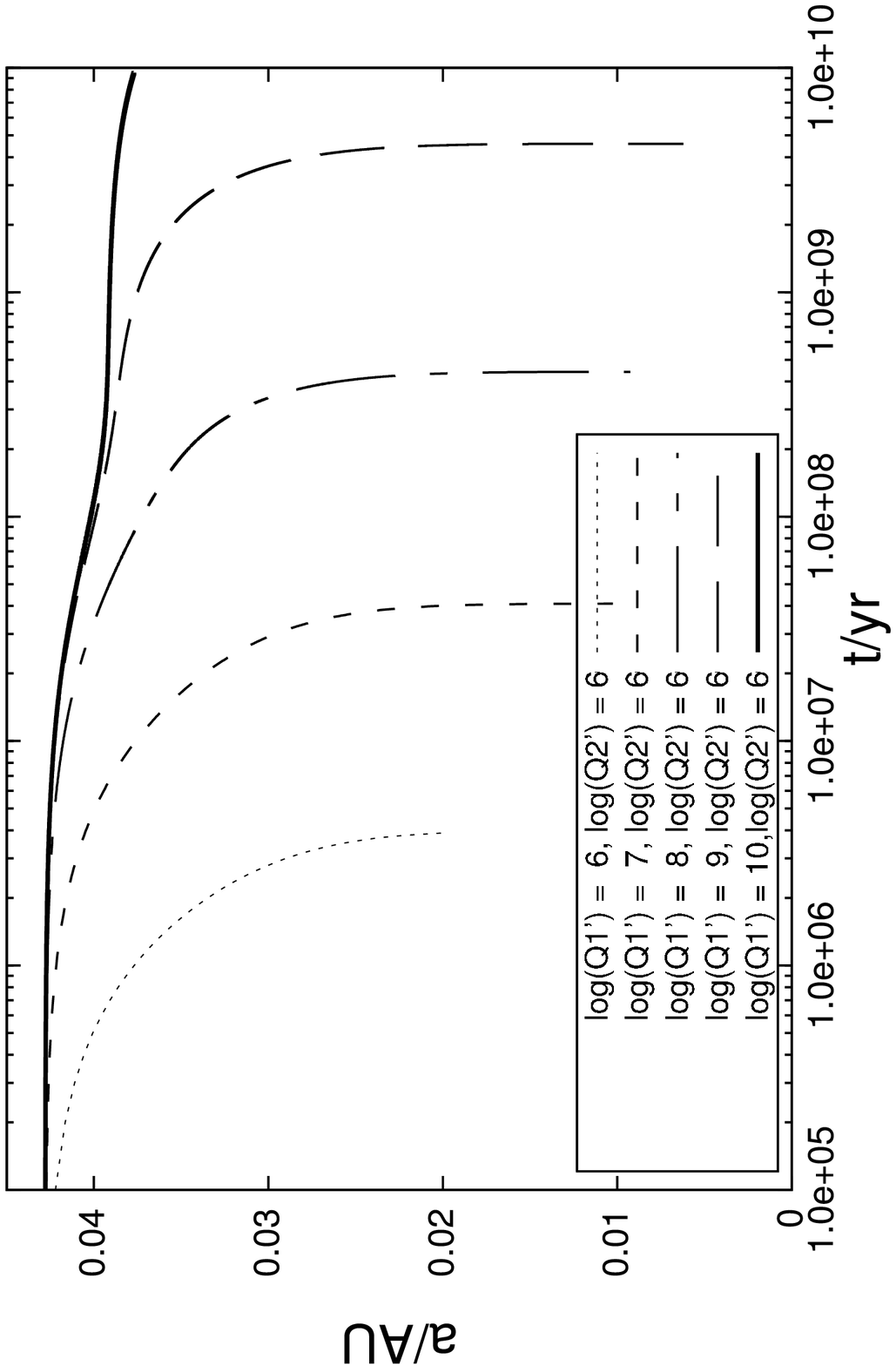}} 
    \subfigure{\label{FigXO-3-2}
      \includegraphics[width=0.47\textwidth]{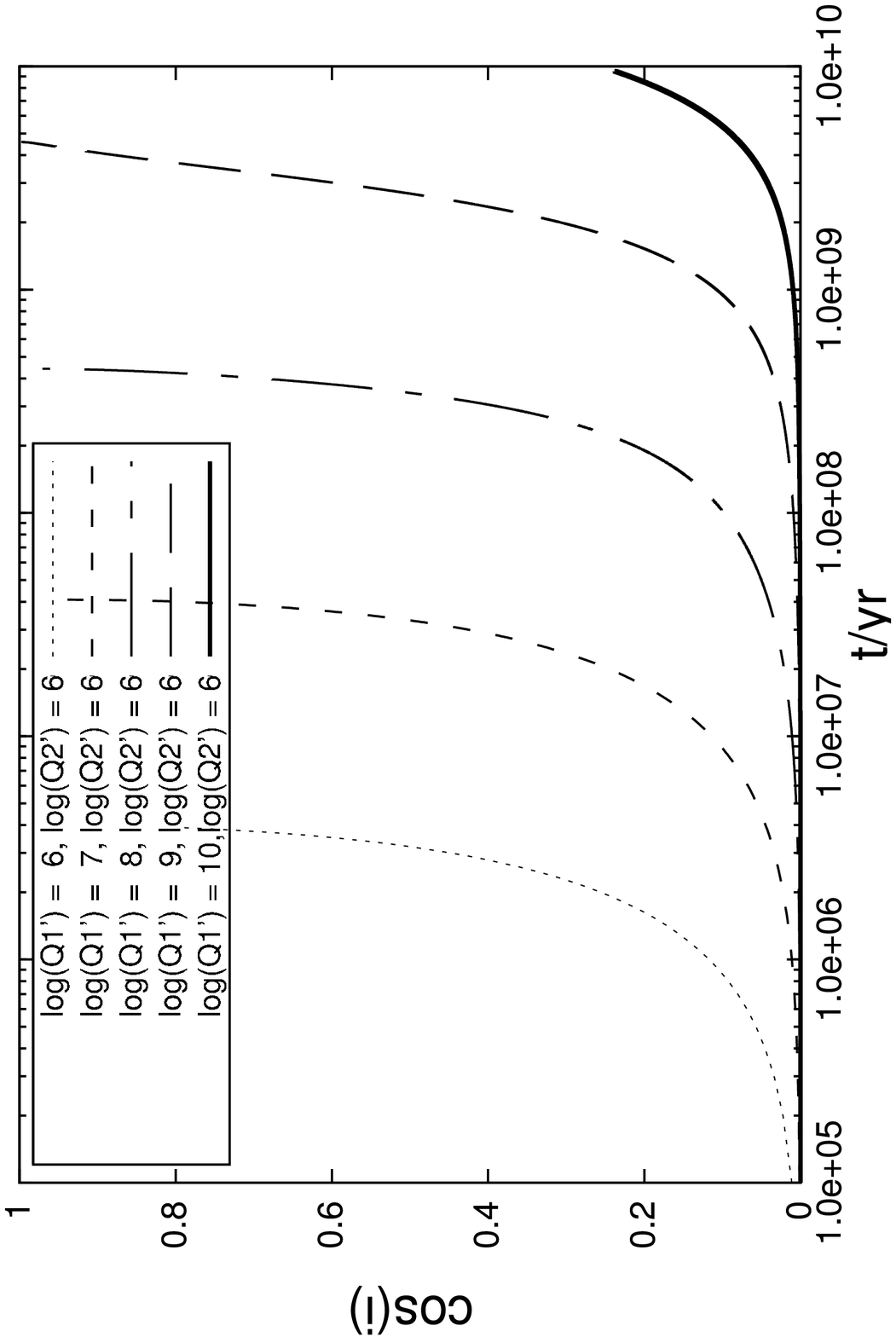}}
    \subfigure{\label{FigXO-3-3}
      \includegraphics[width=0.47\textwidth]{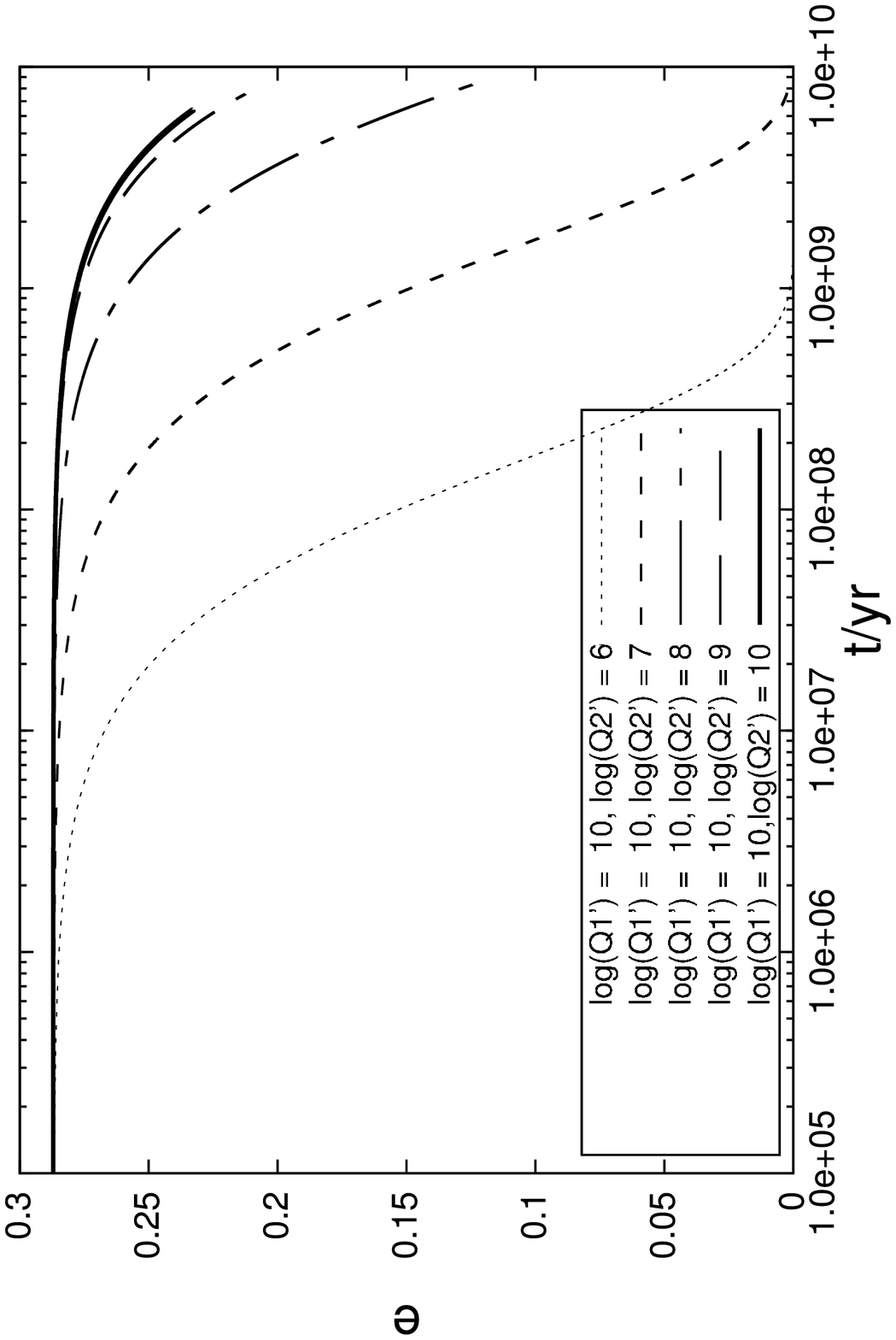}}
  \end{center}
  \caption{Tidal evolution of XO-3 b taking current values for the
    orbital properties of the system, except that $\cos i = 90^{\circ}$ 
    (not unreasonable since this roughly corresponds to the upper
    limit on $\lambda$, which in any case gives a lower bound
    on $i$). Magnetic braking is included with
    $\gamma=0.1$, and $\Omega_{1}/n = 2$ initially (results do not depend
    strongly on this choice). From the top and middle plots we
    require $Q_{1}^{\prime} \geq 10^{10}$ for the planet to survive for
    several Gyr, and maintain its high inclination. From the bottom
    plot we see that
    if $Q_{1}^{\prime} \geq 10^{10}$, we require $Q_{2}^{\prime} \geq 10^{8}$ to
    maintain the current eccentricity for a few Gyr. Tidal dissipation
    in both the planet and star must therefore be weak to explain the current
    configuration of the system.}
  \label{FigXO-3}
\end{figure}

The only system currently observed with a spin--orbit
misalignment\footnote{This spin--orbit misalignment has been confirmed
  since the submission of the present paper by \cite{Winn2009}, who find that the sky--projected
  spin--orbit misalignment angle $\lambda = 37.3^{\circ} \pm 3.7^{\circ}$, which is
  significantly smaller than that found
  by \cite{Hebrard2008}. Nevertheless, our conclusions below should still be valid.} is
XO-3 (\citealt{Hebrard2008};
\citealt{Johns-Krull2008}), which has a sky--projected spin--orbit misalignment angle of
$\lambda \simeq 70^{\circ} \pm 15^{\circ}$. This system has a
very massive $m_{2} = 12.5 \, M_{J}$ planet on a moderately
eccentric $e = 0.29$, $P = 3.2$ d orbit around an F-type star of
mass $m_{1} = 1.3 \, M_{\odot}$. Its age is
estimated to be $\tau_{\star} \simeq (2.4-3.1)$ Gyr. Note that
even if the star is rotating near breakup velocity ($P_{\star}\sim 1$ d), the planet is
still subject to tidal inspiral, since $P_{\star} > P \cos i$ (where we
henceforth assume $i = \lambda$, which may slightly
\textit{underestimate} $i$). If we assume that the angle of
inclination of the stellar equator to the plane of the sky is $\sim
90^{\circ}$, then $P_{\star} = 3.3$ d $\sim P$ i.e. $\Omega \sim n$.

\cite{Hebrard2008} quote a stellar spin--orbit alignment timescale of $\sim 10^{12}$
yr for this system, but we find that this is in error by $\sim
10^{5}$. We believe that the reason for this discrepancy is that 
their estimate was based on
assuming that the spin--orbit alignment time for XO-3 b is the same as for 
HD17156 b (\citealt{Narita2008}; \citealt{Cochran2008}), which is a less massive planet on a much
wider orbit. We find $\tau_{i}\sim 30$ Myr (using the expression in \S
\ref{timescales}) assuming $Q^{\prime}_{1}=10^{6}$ to
align the whole star with the orbit. Circularisation time of
$\tau_{e} \sim 10$ Myr and the inspiral time is estimated to be
$\tau_{a} \sim 16$ Myr from simple estimates.

Integrations for this system are given in Fig.~\ref{FigXO-3} for a variety of
stellar and planetary $Q^{\prime}$ values. These integrations again highlight the
importance of considering coupled evolution
of the orbital and rotational elements, since timescales for tidal
evolution can be quite different from the simple estimates given
above. Indeed, the actual spin--orbit alignment time from integrating the
full set of equations is about an order of magnitude smaller than that
from the simple decay estimate, due to the semi-major axis evolution.

\begin{figure}
\begin{center} 
\subfigure{\includegraphics[width=0.35\textwidth]{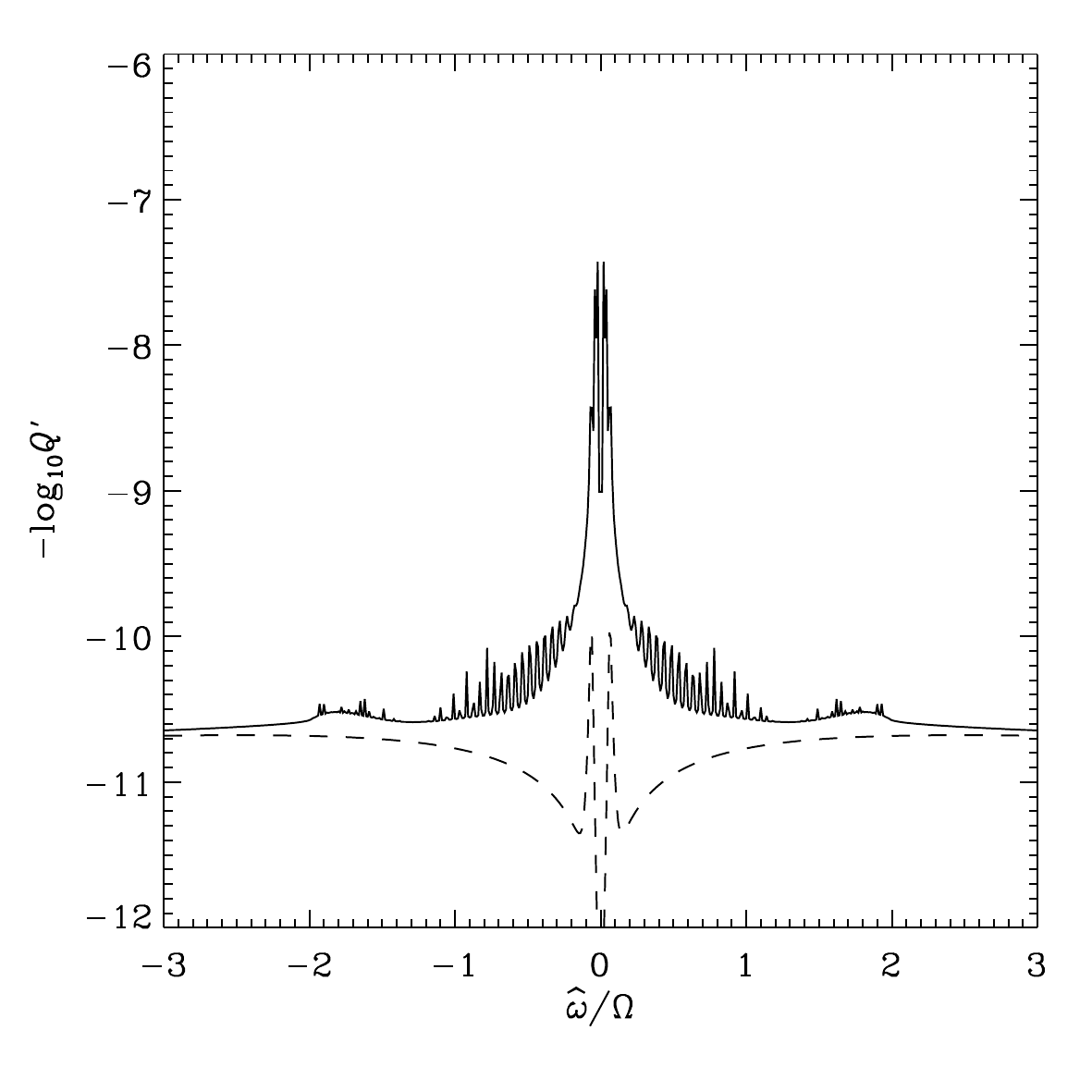}}
\subfigure{\includegraphics[width=0.35\textwidth]{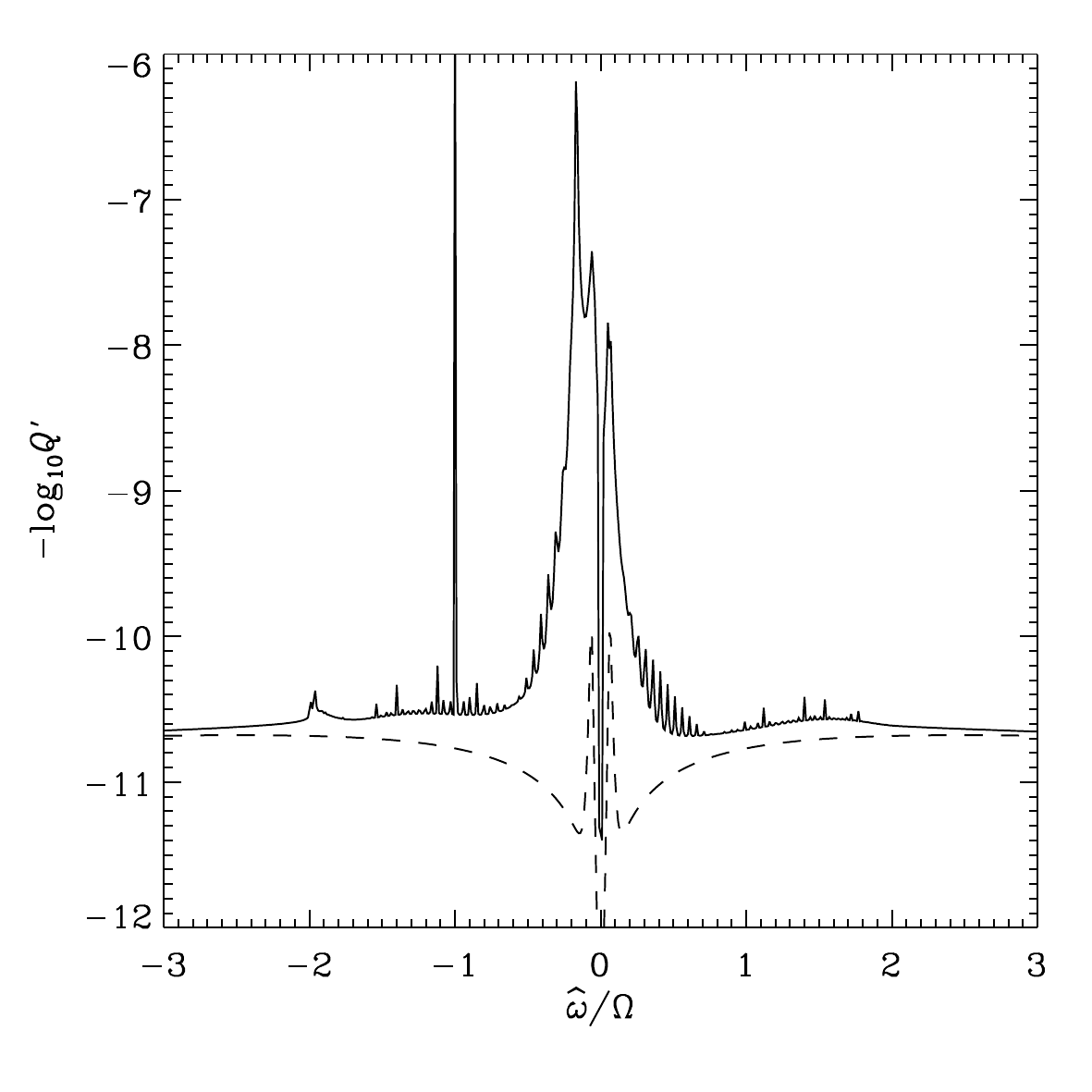}}
\subfigure{\includegraphics[width=0.35\textwidth]{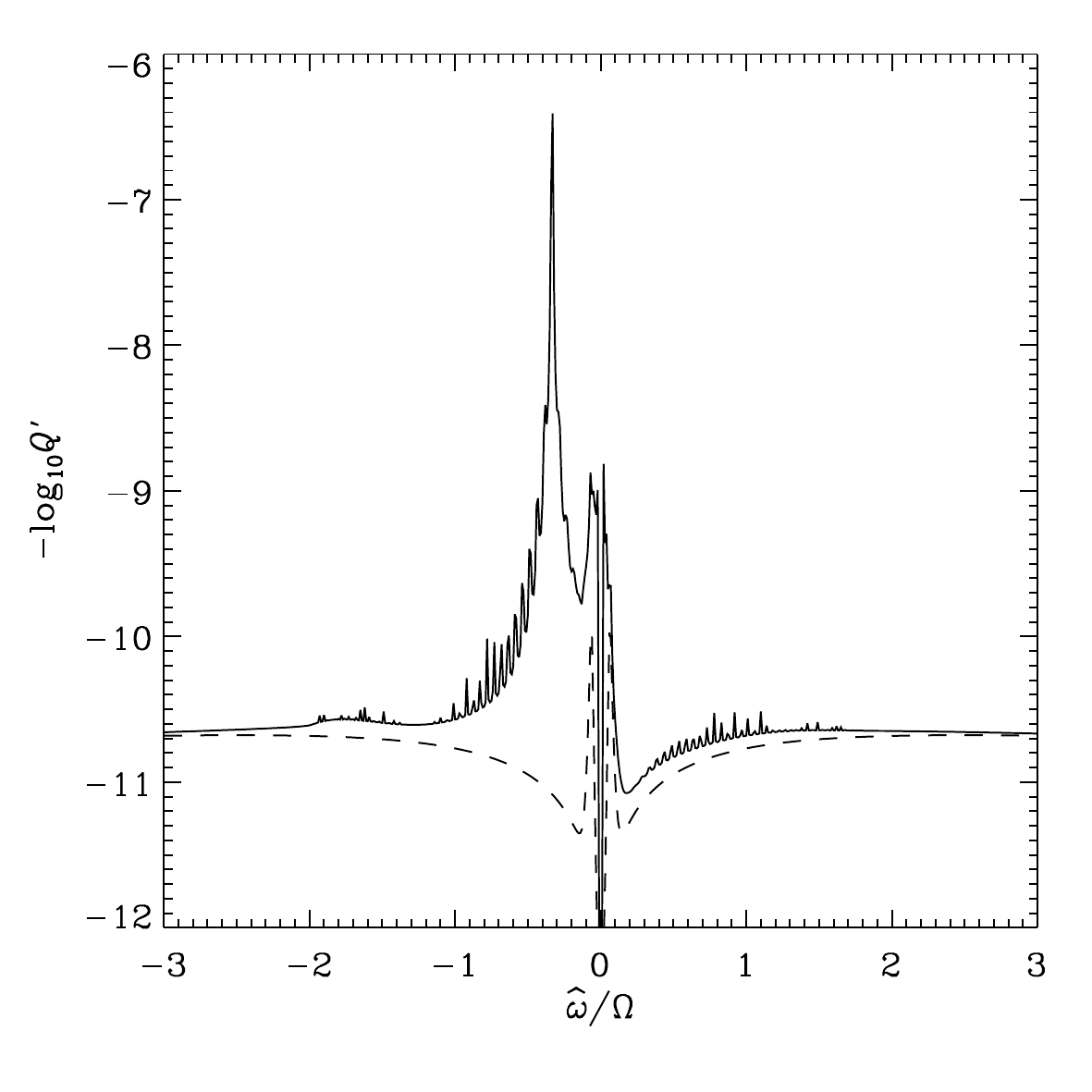}}
\end{center}
\caption{Tidal $Q^{\prime}$-factor as a function of the ratio of tidal
  frequency to spin frequency $\hat \omega/\Omega$, from dissipation
  of the $l = 2$, $m = 0,\, 1,\, 2$ components respectively, of the
  equilibrium tide and dissipation of inertial modes in the OCZ
  of an F-type star (see OL07 for
  details of this calculation). This used a stellar model appropriate
  for XO-3 to model the convection zone. The dashed lines represent
  the effect of omitting the Coriolis force, and therefore inhibiting
  inertial waves. The prominent features in
  each figure (which occur for $\hat \omega/\Omega = -1$ and $-1/6$ for $m=1$ and $\hat
  \omega/\Omega = -1/3$ for $m=2$) are
  Rossby waves, which are probably excited for tidal frequencies not
  relevant for the XO-3 system. For most tidal frequencies $Q^{\prime} \geq
  10^{10}$, which could explain the survival and remnant orbital
  inclination of XO-3 b.}
\label{GioQFstar}
\end{figure}

For the cases considered, the system can only survive and remain with its
current inclination for $\sim 3$ Gyr if
$Q^{\prime}_{1}\geq 10^{10}$. An explanation for the survival and remnant orbital inclination of
XO-3 b could therefore be the inefficiency tidal dissipation in
the host star. The host star is an F--star of mass $m_{1} =
1.3 \pm 0.2 M_{\odot}$, so it will contain a small
convective core and a very thin outer convection zone (OCZ)
separated by a radiative zone. Dissipation in the convective core
will only weakly affect the tide, and
dissipation in the radiation zone will also be weak \citep{Zahn2008}. This is because 
internal inertia--gravity waves excited at the interface between convective and radiative
regions cannot reach the photosphere, where they can damp
efficiently, as supposed for high--mass stars.
In addition, nonlinear effects due to geometrical concentration 
of the waves in the centre of the star cannot occur because 
the waves will reflect from the outer boundary of the inner 
convection zone well before they become nonlinear
(\citealt{GoodmanDickson1998}; OL07).

We expect that most dissipation occurs in the OCZ of the star. A 
calculation of $Q^{\prime}$ for the dissipation of the
equilibrium tide and inertial modes, in the thin OCZ,
using a stellar model\footnote{for which we use EZ Web at
  http://shayol.bartol.udel.edu/$\sim$rhdt/ezweb/}
appropriate for this star (note that the metallicity of the star is
subsolar, with $Z\sim 0.01$), was performed (see
OL07; OL04 for details of the numerical method). 
It must be noted that these calculations involve uncertainties
regarding the effective viscosity of turbulent convection, though 
the general trends in the results below (and in the next section) 
are likely to be quite robust. The
results for the $m = 0,\, 1,\, 2$ components of the tide are plotted in
Fig.~\ref{GioQFstar}. In Appendix \ref{TidalPotential}, we
show that a combination of the $m=0, \, 1, \, 2$ components of the tide
are relevant for spin--orbit alignment and inspiral, and so we must calculate
$Q^{\prime}$ for all components of the $l=2$ (quadrupolar) tide. The
relevant tidal frequencies, assuming $\Omega \sim n$ currently, would be those
of integer $\hat \omega / \Omega$. However, since the angle of 
inclination of the stellar equatorial plane to the plane of the sky
has not been determined, the relevant
tidal frequencies cannot be calculated with any certainty.
Nevertheless, $Q^{\prime} \geq 10^{10}$ for most tidal frequencies for the host star XO-3.
This can explain the survival and remnant inclination of XO-3 b,
since both $\tau_{a}$ and $\tau_{i}$ are now much longer than the 
age of the system. 

In addition, the remnant eccentricity could be maintained due to weak
damping of the tide in the star for the same reasons. 
However, we must also
explain the inefficient damping of the tide in the planet, if indeed the
reason for the eccentricity is that $\tau_{e} >
\tau_{\star}$. 
The planet in this system is massive, and may be a low--mass brown dwarf. If it
formed without a core, then the dissipation of inertial modes may be
reduced if they are able to form global modes, as found in OL04.

An alternative explanation for the survival of XO-3 b involving a larger initial semi-major axis for
  the planet is also possible, but this would require significant tidal
  migration to bring the planet to its current location. This would
require a much lower stellar $Q^{\prime}$ than our calculations
predict, which as discussed in \S \ref{introtide} would imply a very
short inspiral time for the planets on the tightest orbits, if
their host stars have similar $Q^{\prime}$ values. We discuss this
further with regards to the host stars WASP-12 and OGLE-TR-56 in \S \ref{wasp12} below.

\subsection{Tidal dissipation in F--stars}
\label{FstarsQ}

\begin{figure}
  \begin{center} 
    \subfigure{\label{FigQFstar1}
      \includegraphics[width=0.375\textwidth]{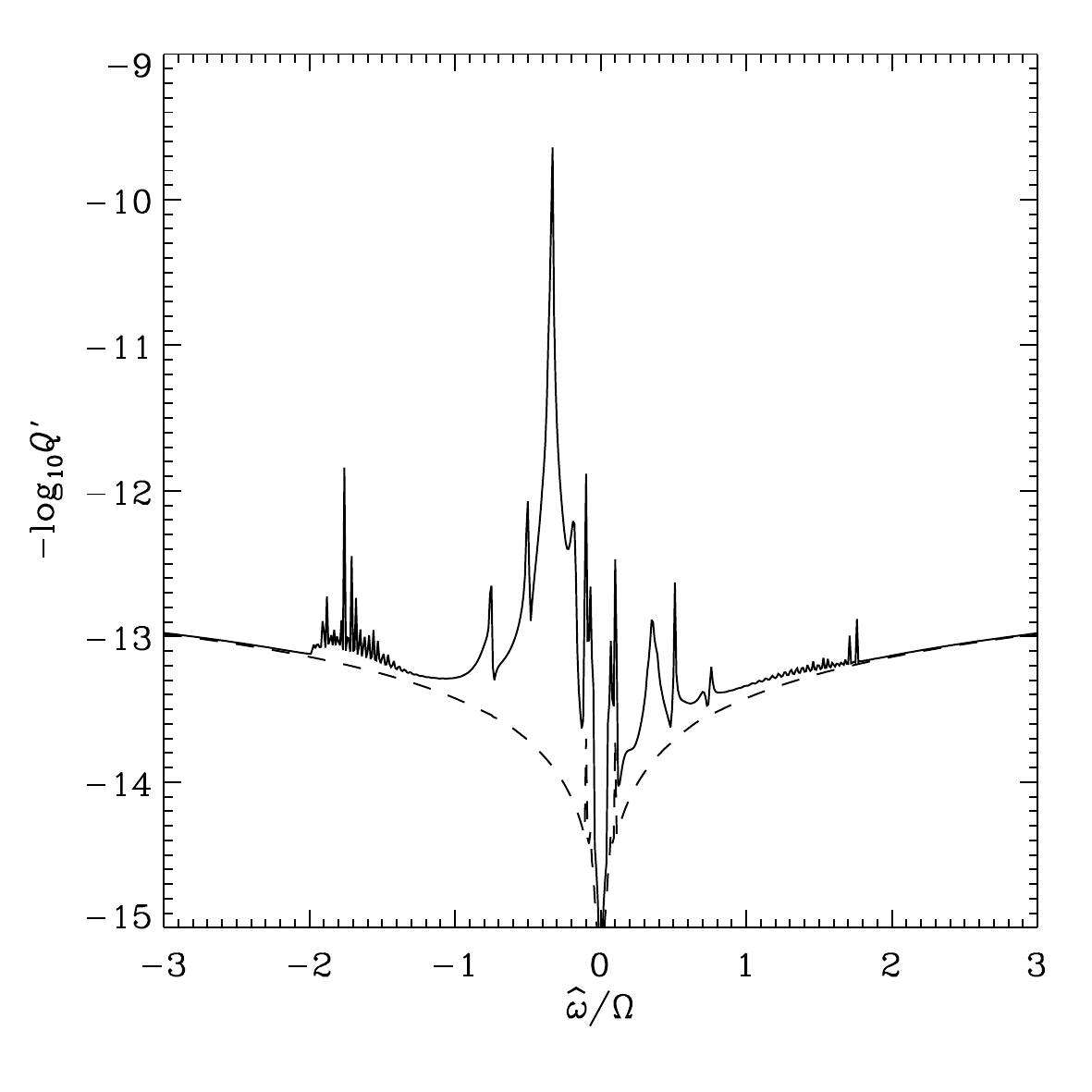}} 
    \subfigure{\label{FigQFstar2}
      \includegraphics[width=0.375\textwidth]{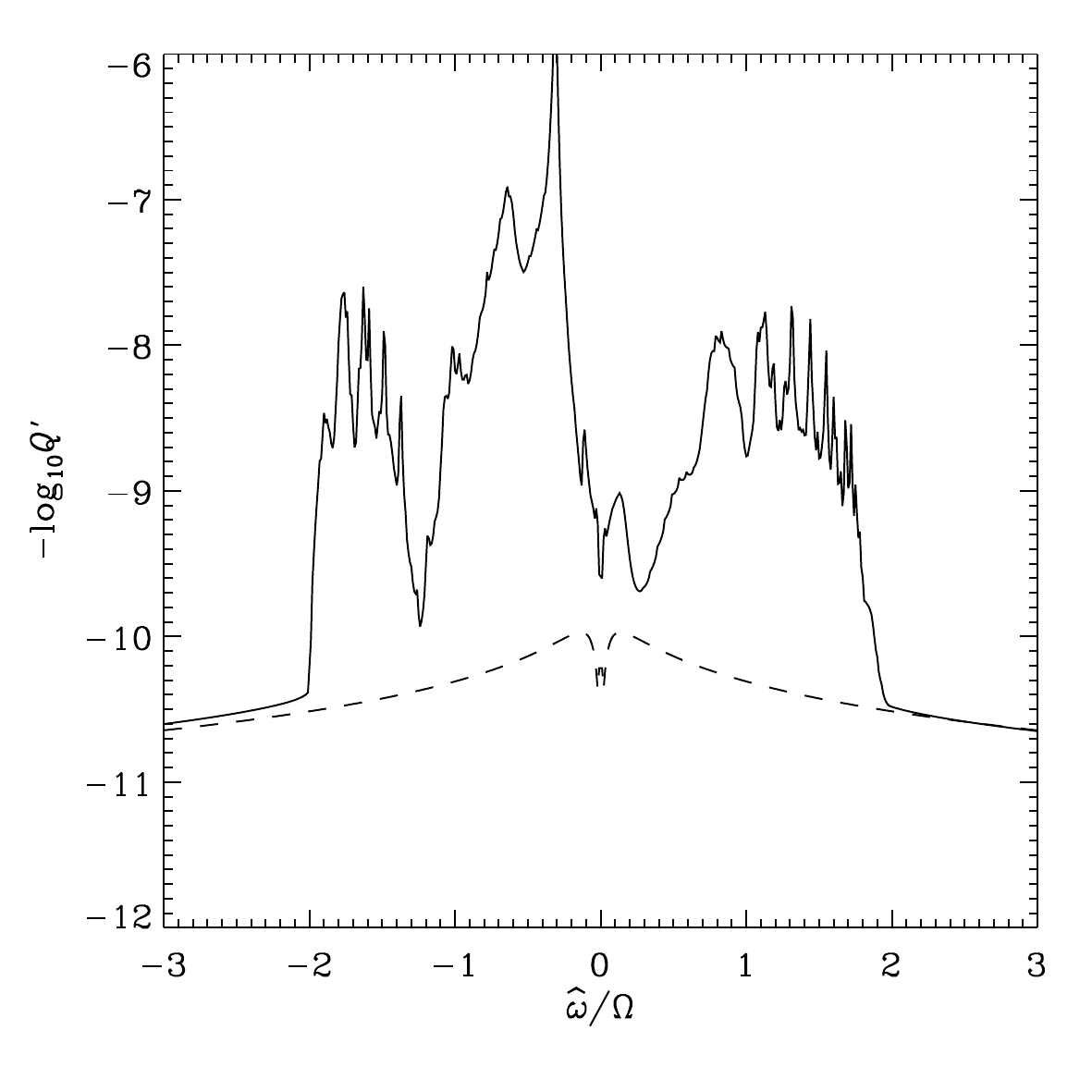}} 
  \end{center}
  \caption{Tidal $Q^{\prime}$-factor as a function of the ratio of tidal
    frequency to spin frequency $\hat \omega/\Omega$, from dissipation
    of the $l=m=2$ component (we find the behaviour of the $m=0$ and 1
    components is similar in magnitude) of the
    equilibrium tide, and dissipation of inertial modes, in the OCZ
    of two F-type star models.  The dashed lines represent
    the effect of omitting the Coriolis force, and therefore inhibiting
    inertial waves. The top figure shows a
    high--mass $1.5 \, M_{\odot}$ F--star, with solar metallicity and an
    age of $0.7$ Gyr, assumed to be spinning with a period of $3$ d. 
    This has a very thin OCZ, and tidal
    dissipation is extremely weak. The bottom figure shows a low--mass 
    $1.2 \, M_{\odot}$ F--star, with solar metallicity and an age of
    $1.0$ Gyr, assumed to be spinning with a period of $3$ d. Tidal
    dissipation is much stronger in this star, though still smaller than
    that in Fig.~6 from OL07 for a $1.0 \, M_{\odot}$ G--star. Together with
    Fig.~\ref{GioQFstar} this shows that
    $Q^{\prime}$ can vary considerably between different stars, even
    within the range of F--stars.}
  \label{FigQFstar}
\end{figure}

Our calculation of $Q^{\prime}$ for XO-3, indicates that it
would be worthwhile to study the range of $Q^{\prime}$ expected for
F--stars, and how these may differ from solar--type stars. We
performed calculations of tidal dissipation
in the OCZs of a variety of F--stars between the
masses of $1.2-1.6 \, M_{\odot}$,
using the numerical method of OL07, and the stellar
models of EZ-Evolution. These stars contain convective cores
surrounded by a radiative zone and an OCZ, and for
the reasons mentioned in the previous section, we expect that tidal
dissipation in the OCZ will dominate the dissipation. 
We consider the range of OCZ properties in F--stars, as the mass
and metallicity are
varied. Below we present a selection of illustrative examples, which
represent the range of properties expected for F--stars. 

Fig.~\ref{FigQFstar} shows $Q^{\prime}$ as a function of tidal
frequency for two F--stars, with different OCZ
properties. Fig.~\ref{FigQFstar1} 
shows that tidal dissipation in more massive F--stars, with very thin OCZs, 
is found to be extremely weak, with $Q^{\prime}\geq
10^{12}$ for most tidal frequencies. This implies that tidal
dissipation in such stars
is probably negligible in contributing 
to the spin--orbit evolution of HJs.
Fig.~\ref{FigQFstar2} shows that tidal dissipation in lower mass
F--stars with thicker convective
envelopes, is similar to but slightly weaker than,
that for solar--type stars (OL07), since $Q^{\prime} \sim 10^{8}$
for most tidal frequencies in the range $|\hat \omega| < 2 |\Omega|$.

A model with $1.3 \, M_{\odot}$ star similar to XO-3, except that we
choose supersolar metallicity ($Z=0.03$), is found to have a similar
OCZ to the $1.2 \, M_{\odot}$ model in
Fig.~\ref{FigQFstar2}, and has a very
similar $Q^{\prime}$. The metallicity of the star affects the 
thickness of the OCZ, and therefore the efficiency of tidal
disspation.

From these examples, it is clear that assuming a single
$Q^{\prime}$ applies for
all stars is probably incorrect. 
Even within the mass range of F--stars there is considerable variation in
$Q^{\prime}\sim 10^{8}-10^{12}$ in our calculations, primarily as a result of the
variation in the mass
fraction contained in the OCZ. 
There are also differences between G and F
stars, due to differences in internal structure -- most notably the
presence of a radiative core in a G--type star may lead to enhanced
dissipation by nonlinear effects. Lower mass
stars, and those with
higher metallicity, tend to have thicker OCZs than higher
mass, low--metallicity stars. In addition, higher mass stars are more
centrally condensed, so the mass fraction in their outer regions will be
lower. This results in low--mass, high--metallicity stars having lower
$Q^{\prime}$ then high--mass, low--metallicity stars. In addition, OLO7
found that the spin period of the star also affects $Q^{\prime}$.

\subsection{The survival of WASP-12 b and OGLE-TR-56 b}
\label{wasp12}

The results of the last two sections also allow us to propose
an explanation for the survival of the planets on the tightest orbits, 
such as WASP-12 b \citep{Hebb2008} and OGLE-TR-56 b
\citep{Sasselov2003}. Taking the current values for the stellar
properties of the host stars in both systems, we find that they
are both likely to have a similar internal
structure to those discussed above, with a convective core present,
which will prevent internal inertia--gravity waves from reaching
the centre of each star. In addition, these stars
are slowly rotating, and so the relevant tidal frequencies are 
likely to be outside the range of inertial waves 
($|\hat \omega| > 2 |\Omega|$), which would imply that such waves
are not excited by tidal forcing. 
This means that $Q^{\prime}$ is likely to be much 
larger than that in the models discussed above 
(in Fig.~\ref{FigQFstar}), particularly since reducing the spin frequency is found to increase
$Q^{\prime}$ for a given ratio $\hat \omega/\Omega$ (OL07). Therefore
it is
likely that the relevant $Q^{\prime}\ga 10^{10}$, which would imply that tidally induced inspiral 
will not occur within the age of the system.
This means that weak dissipation in
the star could potentially explain the survival of both of these planets.

\section{Conclusions}

In this paper we have investigated the long--term tidal evolution of
close--in extrasolar planets. We studied the effects of magnetic braking on tidal
evolution in a simplified system, and then performed numerical
integrations for a variety of HJ systems, with particular emphasis on 
inclination evolution. We now summarise the main results of this work.

Magnetic braking moves the corotation radius of
the star outwards such that any close--in planets will eventually orbit inside
corotation ($\Omega \cos i < n$), and be subject to orbital decay due
to tides. It makes no
sense to refer to any close--in planet as being \textit{tidally evolved},
since magnetic braking removes the only equilibrium state
accessible through tidal friction during the expected stellar lifetime.

Magnetic braking is found to be important for the tidal
evolution of HJs unless $\Omega \cos i \ll n$, in which case the
orbit is already well inside the orbit--projected corotation
radius. Highly inclined (especially retrograde) orbits tend to be affected less by magnetic
braking, since this condition is often satisfied regardless of the
spin rate. Magnetic braking is particularly important when it comes to interpreting the
tidal evolution of observed systems from formation to the present
day, since the star may have been rotating much more rapidly in the
past. Nevertheless, it is probably not a bad approximation to
neglect magnetic braking for calculating the \textit{future} tidal
evolution of most observed HJs, if the star has already spun down so that $\Omega \ll
n$.
 
Combining our results with those of \cite{JacksonI2008}, we find
that coupled evolution of the orbital and rotational elements is
essential to accurately model tidal evolution, and can result in
much faster evolution than simple timescale estimates predict. This is
especially true for highly eccentric orbits, for which the associated
semi-major axis evolution increases the rate of stellar spin--orbit
alignment by up to several
orders of magnitude.

We find that the true timescale for stellar spin--orbit alignment
is comparable to the inspiral time for HJs, therefore the orbits of most close--in
planets have probably not aligned with the spin of the star. Observed inclinations are likely to be a relic
of the migration process. This means that RM observations of
transiting planets can potentially distinguish between migration
caused by planet--planet scattering or Kozai
oscillations combined with tidal dissipation in the star, and that
produced by tidal interaction with the gas disc. 
If the majority of candidates are found with
$\lambda\sim 0$, this strongly disfavours planet--planet scattering or
Kozai migration, since they are expected to produce significantly
inclined orbits, and we have found that tides are unlikely to have aligned orbits
without causing inspiral. Alternatively, if systems are found with
significantly nonzero $\lambda$, 
then some planet--planet scattering or Kozai migration 
could have occurred to produce
these orbital inclinations. We strongly encourage
future observations of the RM effect for transiting planets.

For most HJs, tides tends to circularise the planet's orbit
before spin--orbit alignment or inspiral occurs. If an orbit is initially
inclined and eccentric, as a
result of planet--planet scattering or Kozai migration into a
short--period orbit, then we would
expect the orbit to become circular before
it aligns. Therefore, if we observe a
planet on an inclined, circular orbit, we cannot rule out a
non-negligible eccentricity in the past. This 
means that we should observe fewer eccentric orbits than
inclined orbits
\textit{if those systems start with a uniform distribution of points
  in ($e,i$)--space}, due to tidal friction. This should be considered when
comparing the observed $(e,i)$ distribution with those predicted from
theoretical work on Kozai migration or planet--planet
scattering, before we can further constrain these theories (\citealt{Fabrycky2007}; \citealt{JT2008}).

The misaligned spin and orbit of the XO-3 system could
potentially be explained in terms of inefficient tidal dissipation
inside the host star. The required stellar $Q_{1}^{\prime} \ga 10^{10}$ required
for the survival and remnant misalignment is predicted from
theoretical calculations of tidal dissipation in the OCZ
of an F--star. In addition, the remnant
eccentricity poses constraints on the planetary
$Q_{2}^{\prime} \ga 10^{8}$, in the absence of perturbing forces
that could excite the eccentricity.

The stellar $Q^{\prime}$ has been shown to vary widely between
different stars, and even within the range of F--stars, from
$10^{8}-10^{12}$. There are also
differences between $Q^{\prime}$ for the F--star models discussed here, and the
solar--type star models discussed in OL07. This implies that 
assuming a single value of $Q^{\prime}$ applies to all exoplanet host
stars is probably incorrect. $Q^{\prime}$ has been found to vary with stellar
mass, metallicity, spin period, as well as tidal frequency. 
The presence of a convective core is likely to be important, in that
it acts to prevent internal inertia--gravity waves, which are excited at the interface between the
convective and radiative zones, from reaching the centre of the star,
where nonlinear effects could enhance their dissipation. This may
explain the survival of some of the planets on the tightest orbits, such as 
WASP-12 b and OGLE-TR-56 b, whose host stars are likely to have convective cores.

It is clear that much more work is required to study the mechanisms
of tidal dissipation in rotating stars. It would
also be useful to perform a detailed study into the possibility of 
core--envelope decoupling in such stars, since this may have
implications for the survival
and remnant inclinations of close--in planets. Future observations of
transiting planets will constrain these theories, hopefully leading to a better understanding
of the mechanisms at work in the formation and evolution of planetary systems.

\section*{Acknowledgments}
We would like to thank the referee, Richard Greenberg, for helpful comments
that have improved the manuscript. In addition, A.J.B would like to thank STFC for a research
studentship.

\appendix

\onecolumn

\section[]{Tidal evolution equations}
\label{Appendix}

Here we present the orbit-averaged tidal evolution equations
(\citealt{Eggleton1998}; \citealt{Eggleton2001}; \citealt{ML2002}) used in
this work. These are derived starting with the equation of
relative motion of a planet of mass $m_{2}$ and its host star
of mass $m_{1}$
\begin{eqnarray} 
  \frac{d^{2}\mathbf{r}}{dt^{2}} = - \frac{Gm_{12}}{r^{3}}\mathbf{r} +
  \mathbf{f},
\end{eqnarray}
where $\mathbf{r}$ is the separation, $m_{12} = m_{1}+m_{2}$, and $\mathbf{f}$ represents a
perturbing acceleration. In \cite{Eggleton2001}, $\mathbf{f}$
contains many contributions: the effects of general relativity,
perturbing accelerations due to other planets and tidal and spin
distortions of the star and planet, as well as tidal friction. Here we
solely consider the perturbing effects of tidal friction and set 
$\mathbf{f} = \mathbf{f}_{\mathrm{tf}} = \mathbf{f}^{1}_{\mathrm{tf}}
  + \mathbf{f}^{2}_{\mathrm{tf}}$, where
\begin{eqnarray}
\label{tidalforce}
  \mathbf{f}^{1,2}_{\mathrm{tf}} = - 3\tau_{1,2} k_{1,2} n^{2}
  \left(\frac{m_{2,1}}{m_{1,2}}\right)
  \left(\frac{R_{1,2}}{a}\right)^{5} \left(\frac{a}{r}\right)^{8}
  \left[3(\hat{\mathbf{r}} \cdot \dot{\mathbf{r}})\hat{\mathbf{r}}+(\hat{\mathbf{r}}
      \times \dot{\mathbf{r}}-r\;\mathbf{\Omega}_{1,2})\times \hat{\mathbf{r}}\right]. 
\end{eqnarray}
Here $n = \sqrt{\frac{Gm_{12}}{a^{3}}}$ is the orbital mean motion,
$k_{1,2}$ are the second-order potential Love numbers for the star and
planet respectively (which is twice the apsidal motion constant), 
and $\tau_{1,2}$ is the effective tidal lag time for each body. This form of
the dissipative force of tidal friction
is that derived under the assumption of a constant lag
time in the equilibrium tide model \citep{Eggleton1998}. In this
model, we assume that the body quasi-hydrostatically adjusts to the 
perturbing potential of its companion, but delayed by some small lag time 
($\tau_{1}$ for the star, $\tau_{2}$ for the planet) 
that is proportional to the dissipation. 
Thus for each body, $Q=\frac{1}{\hat \omega \tau}$ is assumed to be
inversely proportional to the tidal frequency $\hat \omega$, so that the lag time $\tau$
is independent of tidal frequency, and is therefore the same for all
components of the tide. This model is beneficial because the resulting secular evolution
equations can treat arbitrary orbital eccentricities
and stellar and planetary obliquities.
 
In the resulting equations we have chosen to parametrise the
efficiency of tidal dissipation in each body by
redefining $Q=\frac{1}{n\tau}$, and adopt a constant $Q$
which does not change during the evolution i.e. we assume effectively
that the lag time scales with the orbital period (then
Eq.~\ref{tidalforce} matches Eq.~4 in \citealt{ML2002}, where this
assumption was not made explicit). We also introduce
the definition $Q^{\prime} = \frac{3Q}{2k}$, as discussed in
\S \ref{introtide}. This allows us to
discuss ``$Q^{\prime}$ values'' for particular bodies, which do not
change as the orbital and rotational elements vary. For the purposes
of this paper we \textit{define} $Q^{\prime} =
\frac{3}{2kn\tau}$. This is equivalent to \textit{assuming} that the
relevant tidal frequency $\hat \omega = n$. Note that this may not
give identical numerical factors in the resulting equations to other formulations of tidal
friction (e.g. \citealt{GoldSot1966}; \citealt{Zahn1977};
\citealt{Hut1981}), but we feel that this is the best way to study the general
effects of tidal friction, given our uncertainties in the value of $Q^{\prime}$,
and its dependence on $\hat \omega$, for realistic giant
planets and stars.

The secular evolution of the orbital elements is calculated via the
specific orbital angular momentum vector $\mathbf{h}= \mathbf{r} \times
  \dot{\mathbf{r}} = n a^{2}
\sqrt{1-e^{2}} \; \hat{\mathbf{h}}$, and the
eccentricity
vector $\mathbf{e}$. The eccentricity vector has the
magnitude of the eccentricity, and points in the direction of
periastron, and is given by
\begin{eqnarray}
 \mathbf{e} = \frac{\dot{\mathbf{r}} \times \mathbf{h}}{Gm_{12}} - \hat{\mathbf{
    r}}. 
\end{eqnarray}
The planet's specific orbital angular momentum $\mathbf{h}$ changes at
a rate
\begin{eqnarray} 
  \label{B1}
  \frac{d\mathbf{h}}{dt} = \mathbf{r}\times
  (\mathbf{f}^{1}_{\mathrm{tf}}
  +\mathbf{f}^{2}_{\mathrm{tf}}),
\end{eqnarray}
with a corresponding rate of angular momentum transfer between the
orbit and spin of each body given by
\begin{eqnarray} 
  \dot{\mathbf{J}}_{1,2} = I_{1,2}\dot{\mathbf{\Omega}}_{1,2} = -\mu
  \mathbf{r \times f}^{1,2}_{\mathrm{tf}}, \end{eqnarray}
since total angular momentum is conserved, and where $\mu=\frac{m_{1}m_{2}}{m_{12}}$ is the reduced mass of the
system. The eccentricity vector evolves as
\begin{eqnarray} 
  \label{B2}
  \frac{d\mathbf{e}}{dt} = \frac{\left[2(\mathbf{f}_{\mathrm{tf}}\cdot \dot{\mathbf{r}})\mathbf{r} -
      (\mathbf{r} \cdot \dot{\mathbf{r}})\mathbf{f}_{\mathrm{tf}} -
      (\mathbf{f}_{\mathrm{tf}}\cdot \mathbf{
	r}) \dot{\mathbf{r}} \right]}{Gm_{12}}.
\end{eqnarray}
Eqs.~\ref{B1} and \ref{B2} are time-averaged over the orbit of the planet, and
the resulting differential equations are given below. Numerical
integration of these equations gives 
the secular evolution of the orbital elements. Note that we have written 
these equations so that they are regular at $e=0$, unlike
those in \cite{Eggleton2001} and \cite{ML2002} -- we
eliminate reference to the basis vectors chosen in their
representation, since $\hat{\mathbf{e}}$ is undefined for a
circular orbit, whereas $\mathbf{e}$ is well defined and
equal to zero.

 \begin{eqnarray}
    \frac{d\mathbf{h}}{dt} &=& -\frac{1}{t_{f1}}\left[\frac{\mathbf{\Omega}_{1} \cdot \mathbf{e}}{2n}f_{5}(e^{2})h\mathbf{{e}} 
      - \frac{\mathbf{\Omega}_{1}}{2n}f_{3}(e^{2})h
      +\left(f_{4}(e^{2})-\frac{\mathbf{\Omega}_{1} \cdot \mathbf{
	  h}}{2n}\frac{1}{h}f_{2}(e^{2})\right)\mathbf{h}\right] \\
    &&-\frac{1}{t_{f2}}\left[\frac{\mathbf{\Omega}_{2} \cdot \mathbf{e}}{2n}f_{5}(e^{2})h\mathbf{{e}} 
      - \frac{\mathbf{\Omega}_{2}}{2n}f_{3}(e^{2})h
      +\left(f_{4}(e^{2})-\frac{\mathbf{\Omega}_{2}\cdot \mathbf{
	  h}}{2n}\frac{1}{h}f_{2}(e^{2})\right)\mathbf{h}\right] \\
    &=&
    \left(\frac{d\mathbf{h}}{dt}\right)_{1}+\left(\frac{d\mathbf{h}}{dt}\right)_{2}
    \\
    h\frac{d\mathbf{e}}{dt} &=& -\frac{1}{t_{f1}}\left[\frac{\mathbf{\Omega}_{1}
	  \cdot \mathbf{e}}{2n} f_{2}(e^{2})\mathbf{h}
      +9\left(f_{1}(e^{2})h - \frac{11}{18}\frac{\mathbf{\Omega}_{1} 
	\cdot \mathbf{h}}{n}
      f_{2}(e^{2})\right)\mathbf{e}\right] \\
    && -\frac{1}{t_{f2}}\left[\frac{\mathbf{\Omega}_{2}
	  \cdot \mathbf{e}}{2n} f_{2}(e^{2})\mathbf{h}
      +9\left(f_{1}(e^{2})h - \frac{11}{18}\frac{\mathbf{\Omega}_{2}
    \cdot \mathbf{h}}{n}
      f_{2}(e^{2})\right)\mathbf{e}\right]
    \\ \frac{d\mathbf{\Omega}_{1}}{dt} &=&
    -\frac{\mu}{I_{1}}\left(\frac{d\mathbf{h}}{dt}\right)_{1}
    +\dot {\bmath{\omega}}_{\mathrm{mb}} \\ &=& \frac{\mu}{I_{1}
      t_{f1}}\left[\frac{\mathbf{\Omega}_{1} 
	  \cdot \mathbf{e}}{2n}f_{5}(e^{2})h\mathbf{{e}} 
      - \frac{\mathbf{\Omega}_{1}}{2n}f_{3}(e^{2})h
      +\left(f_{4}(e^{2})-\frac{\mathbf{\Omega}_{1} \cdot \mathbf{
	  h}}{2n}\frac{1}{h}f_{2}(e^{2})\right)\mathbf{h}\right]+\dot {\bmath{\omega}}_{\mathrm{mb}}
    \\ \frac{d\mathbf{\Omega}_{2}}{dt} &=&
    -\frac{\mu}{I_{2}}\left(\frac{d\mathbf{h}}{dt}\right)_{2} \\ &=&
    \frac{\mu}{I_{2} t_{f2}}
    \left[\frac{\mathbf{\Omega}_{2}\cdot \mathbf{e}}{2n}f_{5}(e^{2})h\mathbf{e} 
      - \frac{\mathbf{\Omega}_{2}}{2n}f_{3}(e^{2})h
      +\left(f_{4}(e^{2})-\frac{\mathbf{\Omega}_{2} \cdot \mathbf{
	  h}}{2n}\frac{1}{h}f_{2}(e^{2})\right)\mathbf{h}\right]
\end{eqnarray} 

We also need to define an inverse tidal friction timescale
$t^{-1}_{f}$ for each body (here for body 1, change $1\rightarrow2$
to get the corresponding expression for body 2), and the functions of the eccentricity
(first derived in a similar form by \citealt{Hut1981}).
\allowdisplaybreaks{
 \begin{eqnarray}
    \frac{1}{t_{f1}} &=&
    \left(\frac{9n}{2Q^{\prime}_{1}}\right)\left(\frac{m_{2}}{m_{1}}\right)\left(\frac{R_{1}}{a}\right)^{5} =
    \sqrt{Gm_{12}}\left(\frac{9}{2Q^{\prime}_{1}}\right)\left(\frac{m_{2}}{m_{1}}\right)R_{1}^{5}\;a^{-\frac{13}{2}} \\
    f_{1}(e^{2}) &=&
    \frac{1+\frac{15}{4}e^{2}+\frac{15}{8}e^{4}+\frac{5}{64}e^{6}}{(1-e^{2})^{\frac{13}{2}}}
    \\
    f_{2}(e^{2}) &=&
    \frac{1+\frac{3}{2}e^{2}+\frac{1}{8}e^{4}}{(1-e^{2})^{5}} \\
    f_{3}(e^{2}) &=&
    \frac{1+\frac{9}{2}e^{2}+\frac{5}{8}e^{4}}{(1-e^{2})^{5}} \\
    f_{4}(e^{2}) &=&
    \frac{1+\frac{15}{2}e^{2}+\frac{45}{8}e^{4}+\frac{5}{16}e^{6}}{(1-e^{2})^{\frac{13}{2}}} \\
    f_{5}(e^{2}) &=& \frac{3+\frac{1}{2}e^{2}}{(1-e^{2})^{5}} \\
    f_{6}(e^{2}) &=&
    \frac{1+\frac{31}{2}e^{2}+\frac{255}{8}e^{4}+\frac{185}{16}e^{6}+\frac{25}{64}e^{8}}{(1-e^{2})^{8}} \\
\end{eqnarray} 
}
In the absence of magnetic braking ($\dot {\bmath{\omega}}_{\mathrm{mb}}$) the total angular momentum is
conserved ($\frac{d\mathbf{L}}{dt}=\mathbf{0}$). With the inclusion of magnetic braking, the total
angular momentum of the system decreases at a rate
\begin{eqnarray}
\frac{d\mathbf{L}}{dt} = \frac{d}{dt}\left(\mu\mathbf{h} + I_{1}\mathbf{\Omega}_{1} +
I_{2}\mathbf{\Omega}_{2}\right) = I_{1}\dot {\bmath{\omega}}_{\mathrm{mb}} =
-\alpha_{\mathrm{mb}} I_{1} \Omega_{1}^{2} \mathbf{\Omega}_{1}.
\end{eqnarray}

Tidal dissipation and magnetic braking result in a loss of energy from the system. The
total rate of energy dissipated can be expressed as follows:
\begin{eqnarray}
  \dot E_{tot} &=& \frac{1}{2}\frac{Gm_{1}m_{2}}{a}\frac{\dot a}{a} +
  I_{1}\Omega_{1}\dot \Omega_{1} + I_{2}\Omega_{2}\dot \Omega_{2} \\ 
  &=& H_{1} + H_{2} -\alpha_{\mathrm{mb}} I_{1}\Omega_{1}^{4},
\end{eqnarray}
where the final term in the last expression is the energy dissipated
by magnetic braking and
\begin{eqnarray}
  H_{1} &=& -\frac{\mu
    h}{n t_{f1}}\left[\frac{1}{2}\left(\Omega_{1}^{2}f_{3}(e^{2})+\frac{(\mathbf{\Omega}_{1}\cdot
	\mathbf{h})^{2}}{h^{2}}f_{2}(e^{2}) - (\mathbf{\Omega}_{1}\cdot \mathbf{
      e})^{2}f_{5}(e^{2})\right) - 2n\frac{(\mathbf{\Omega}_{1}\cdot \mathbf{
	h})}{h}f_{4}(e^{2})
    +n^{2} f_{6}(e^{2})\right], \\
  H_{2} &=& -\frac{\mu
    h}{n
    t_{f2}}\left[\frac{1}{2}\left(\Omega_{2}^{2}f_{3}(e^{2})+\frac{(\mathbf{\Omega}_{2}\cdot
    \mathbf{
	h})^{2}}{h^{2}}f_{2}(e^{2}) - (\mathbf{\Omega}_{2}\cdot \mathbf{
      e})^{2}f_{5}(e^{2})\right) - 2n\frac{(\mathbf{\Omega}_{2}\cdot \mathbf{
	h})}{h}f_{4}(e^{2})
    +n^{2} f_{6}(e^{2})\right],
\end{eqnarray}
are the tidal heating rates in the star and planet, respectively. It has
been proposed that $H_{2}$ can result in planetary inflation, if the
energy is deposited deep in the interior of the planet (\citealt{BLM2001};
\citealt{JacksonII2008}). Note that
these expressions approach zero as $e\rightarrow 0$,
$\mathbf{\Omega}_{1,2}\cdot \mathbf{e}\rightarrow 0$
and $\mathbf{\Omega}_{1,2}\cdot \mathbf{h} \rightarrow n$ i.e. tidal
dissipation continues until the orbit becomes circular, and the spin
of each body becomes coplanar and synchronous with the orbit.

\section[]{Tidal potential expanded to arbitrary order in stellar obliquity}
\label{TidalPotential}

In this section we expand the tidal potential into its separate
Fourier components, following the approach of OL04. 
We aim to study which tidal frequencies are
relevant for highly inclined orbits, such as that of XO-3 b. 
We consider two bodies in mutual Keplerian orbit with semi-major axis
$a$ and mean motion $n$. Adopt a coordinate
system with origin at the centre of body 1, and let this represent the
star, and body 2 the planet. The tidal potential experienced at an
arbitrary point P in body 1 is the nontrivial
term of lowest order in $r$,
\begin{eqnarray}
\Psi = \frac{Gm_{2}}{2R^{5}}\left[R^{2}r^{2}-3(\mathbf{R} \cdot \mathbf{r})^{2}\right],
\end{eqnarray}
where position vector of the point P in body 1 is $\mathbf{r}$, and
the position vector of the centre of mass of body 2 is
$\mathbf{R}(t)$. Body 2 is treated as a point mass, of mass $m_{2}$.
We consider an inclined, circular orbit. Without loss of generality, 
we consider body 2 to orbit in a plane inclined to the $(x,y)$-plane by
an angle $i$, so that its Cartesian coordinates are
\begin{eqnarray}
\mathbf{R} = a\left(\cos nt \cos i , \sin nt , \cos nt \sin i \right),
\end{eqnarray}
while the Cartesian coordinates of a point P are
\begin{eqnarray}
\mathbf{r} = r\left( \sin \theta \cos \phi , \sin \theta \sin \phi
, \cos \theta \right),
\end{eqnarray}
where $\left(r,\theta ,\phi \right)$ are the usual spherical polar
coordinates. Then
\begin{eqnarray}
\Psi = \frac{Gm_{2}}{2a^{3}}r^{2}\left[1 - 3\left(\cos nt \cos i
  \sin \theta \cos \phi + \sin nt \sin \theta \sin \phi + \cos nt \sin
  i \cos \theta \right)^{2}\right] .
\end{eqnarray}
Let 
\begin{eqnarray}
\tilde P^{m}_{l}(\cos\theta ) =
\left[\frac{(2l+1)(l-m)!}{2(l+m)!}\right]^{\frac{1}{2}}P^{m}_{l}(\cos
\theta ),
\end{eqnarray}
where $0 \leq m \leq l$ and $l \in \mathbb{Z^{+}}$, denote an associated Legendre polynomial
normalised such that 
\begin{eqnarray}
\int_{0}^{\pi}\left[\tilde P^{m}_{l}(\cos \theta )\right]^{2} \sin
\theta d \theta = 1.
\end{eqnarray}
The tidal potential correct to arbitrary order in the stellar
obliquity $i$ can be expanded as a series of rigidly rotating solid
spherical harmonics of second degree,
\begin{eqnarray}
\Psi = \frac{Gm_{2}}{a^{3}} \left[\sum_{j=1}^{8} A_{j}(i)
  r^{2} \tilde P^{m_{j}}_{2}(\cos \theta ) \cos(m_{j}\phi-\omega_{j}t) \right],
\end{eqnarray}
where the azimuthal order $m_{j}$, frequency $\omega_{j}$,
tidal (Doppler-shifted) forcing frequency $\hat \omega_{j}=\omega_{j}
- m_{j}\Omega$, and the obliquity dependent amplitude $A_{j}(i)$ of
each component are
\begin{flushleft}
\begin{eqnarray}
m_{1} &=& 0, \;\;\;\; \omega_{1} = 0, \;\;\;\;\; \hat \omega_{1} = 0,
\;\;\;\;A_{1} = \sqrt{\frac{1}{10}} \left(1 -\frac{3}{2}\sin^{2} i
\right), \\
m_{2} &=& 2, \;\;\;\; \omega_{2} = 2n, \;\;\;\;\; \hat \omega_{2} = 2n-2\Omega,
\;\;\;\;A_{2} = -\sqrt{\frac{3}{5}} \cos^{4} \frac{i}{2}, \\
m_{3} &=& 0, \;\;\;\; \omega_{3} = 2n, \;\;\;\;\; \hat \omega_{3} = 2n,
\;\;\;\;A_{3} = -\frac{3}{2}\sqrt{\frac{1}{10}}\sin^{2} i , \\
m_{4} &=& 1, \;\;\;\; \omega_{4} = 0, \;\;\;\;\; \hat \omega_{4} = -\Omega,
\;\;\;\;A_{4} = \sqrt{\frac{3}{5}} \cos i \sin i ,\\
m_{5} &=& 1, \;\;\;\; \omega_{5} = 2n, \;\;\;\;\; \hat \omega_{5} = 2n-\Omega,
\;\;\;\;A_{5} = \frac{1}{2}\sqrt{\frac{3}{5}} \sin i\left( \cos i +1 \right), \\
m_{6} &=& 1, \;\;\;\; \omega_{6} = -2n, \;\;\;\;\; \hat \omega_{6} = -2n-\Omega,
\;\;\;\;A_{6} = \frac{1}{2}\sqrt{\frac{3}{5}} \sin i\left( \cos i -1 \right), \\
m_{7} &=& 2, \;\;\;\; \omega_{7} = 0, \;\;\;\;\; \hat \omega_{7} = -2\Omega,
\;\;\;\;A_{7} = \frac{1}{2}\sqrt{\frac{3}{5}} \sin^{2} i , \\
m_{8} &=& 2, \;\;\;\; \omega_{8} = -2n, \;\;\;\;\; \hat \omega_{8} = -2n-2\Omega,
\;\;\;\;A_{8} = -\sqrt{\frac{3}{5}} \sin^{4} \frac{i}{2}.
\end{eqnarray} 
\end{flushleft}
There are eight components of the tide that contribute for
arbitrary stellar obliquity. For small inclination, the terms that are
of first order in the obliquity are for $m=1$ and are the $j=4$ and
$j=5$ components above (not the component having tidal frequency $\hat \omega = n - \Omega$, as incorrectly
stated in OL04). For a coplanar orbit, only $j=1$ and $j=2$ components
above contribute, and these reduce to the first two components in OL04.

For an orbit at $i\sim90^{\circ}$, all components except $j=2 $ and $j=4$
contribute to the tidal force. Thus, for XO-3, it is important to calculate the
resulting $Q^{\prime}$ for the $m=0$, $m=1$ and $m=2$ components of
the tide. The relevant tidal frequencies for this system cannot be
calculated with any certainty, since the stellar spin period has not
been accurately determined. If we assume that the angle of inclination
of the stellar equatorial plane to the plane of the sky is $\sim
90^{\circ}$, then we have $\Omega \sim n$ currently. The relevant
tidal frequencies would then be $\hat \omega = 0,\pm \Omega, \pm 2\Omega,
\pm 3\Omega, \pm 4\Omega$.

\twocolumn

\newcommand{\bibfont}{\small}
\setlength{\bibsep}{0pt}
\bibliography{tidbib}
\bibliographystyle{mn2e}
\end{document}